\documentclass[]{JINST}
\usepackage{graphicx}
\pdfoutput=1
\usepackage{amssymb}
\usepackage{times}
\usepackage{mathptmx}
\usepackage{float}

\usepackage[T1]{fontenc}
\usepackage[latin1]{inputenc}

\providecommand{\tabularnewline}{\\}

\makeatletter

\title{Study of the electron trigger efficiency \\of the CMS Experiment
\\using test beam data}

\author{P. Q. Ribeiro$^a$, M. Gallinaro$^a$ and J. Varela$^{a,b}$
\\\llap{$^a$}LIP,\\Av. Elias Garcia 14, 1000-149 Lisboa, Portugal\\
\\\llap{$^b$}CERN,\\CH-1211 Geneve 23, Switzerland\\
\\Email: \email{ribeiro@lip.pt}}

\abstract{
A study of the electron identification and selection efficiency of
the L1 Trigger algorithm has been performed using the combined ECAL/HCAL
test beam data. A detailed discussion of the electron isolation and
its impact on the selection efficiency is presented. The L1 electron
algorithm is studied for different beam energies and the results indicate
that efficiencies of 98\% or more can be achieved for electrons with
energies between 15 and 100 GeV. The fraction of charged hadrons with
energies from 3 up to 100 GeV rejected by the L1 electron trigger
algorithm is estimated to be larger than 93\%. 
}

\keywords{Calorimeters;  
Trigger concepts and systems (hardware and software)}

\makeatother

\begin{document}

\section{Introduction}

The CMS Level 1 (L1) trigger aims at reducing the event collection
rate to a maximum of 100kHz set by the High Level Trigger (HLT) processing
capacity, while keeping high efficiency for potentially interesting
physics objects. The calorimeter system provides triggers based upon
the energy profile deposited in the CMS calorimeters by objects such
as electrons, photons and jets. The design of the L1 calorimeter algorithms
was based on fast detector simulation and validated using full detector
simulation \cite{CMS_Note_98-027}. In this note, a first study is
presented concerning the estimate of the selection efficiency of the
L1 trigger algorithm for electrons using combined ECAL/HCAL test beam
data. 

This note is organized as follows: Section \ref{sec:L1-electron-Trigger}
contains a brief overview of the conceptual design and implementation
of the L1 electron trigger algorithm and Section \ref{sec:Experimental-Setup}
describes the experimental setup and the data selection. In Section
\ref{sec:Impact-of-Combined} the impact of the isolation criterion
on the electron efficiency is addressed. Results on electron selection
efficiency are presented in Section \ref{sec:Results-on-Electron},
while the results on charged hadron rejection of the L1 electron/photon
algorithm are presented in Section \ref{sec:Results-on-Rejection}.
Finally, conclusions are given in Section \ref{sec:Conclusions}.

\section{L1 Electron Trigger Algorithm\label{sec:L1-electron-Trigger}}

\subsection{Overview of Conceptual Design \label{sub:Overview-of-Conceptual}}

The CMS electromagnetic calorimeter (ECAL) is made of ${\rm PbWO_{4}}$
crystals, with a radiation length $X_{0}$ of 0.89 cm and a Moliere
radius of 2.19 cm. The ECAL is composed of a Barrel covering $\left|\eta\right|\leq1.479$
and two Endcaps covering $1.479\leq\left|\eta\right|\leq3.0$. The
Barrel is divided in two halves, each made of 18 supermodules containing
1700 crystals each. The lateral transverse size of the Barrel crystals
varies slightly according to the rapidity position. The Endcaps consist
of two detectors, a preshower detector (ES) followed by ${\rm PbWO_{4}}$
calorimetry. Each Endcap is divided in two halves, or {}``Dees'',
and consists of 7324 identically shaped crystals grouped in mechanical
units of $5\times5$ crystals (supercrystals). The Barrel and Endcap
crystals are mounted in a quasi-projective geometry so that their
axes have a small angle (\textasciitilde{} 3º) with respect to the
vector from the nominal interaction vertex, in both the $\eta$ and
$\phi$ projections. The CMS hadronic calorimeter (HCAL) is organized
in four subsystems: Barrel (HB), Endcap (HE), Outer (HO) and Forward
(HF). The HB and HE are joined hermetically, surround completely the
ECAL and are also mounted on a quasi projective-geometry. The HB is
an assembly of two half Barrels, each composed of 18 identical $20º$
wedges in $\phi$. Each wedge is composed of flat brass alloy absorber
plates. The innermost and outermost absorber layers are made of stainless
steel. There are 17 active plastic scintillator tiles interspersed
between the stainless steel and brass absorber plates. The HE is composed
entirely of brass absorber plates in a 18-fold $\phi$ geometry matching
that of the Barrel calorimeter. In the HE there are 19 active plastic
scintillator layers.

The L1 electron/photon algorithm uses a $3\times3$ trigger tower
sliding window technique to find electron/photon candidates in the
full $\eta\times\phi$ region covered by the CMS electromagnetic calorimeter.
The sliding window is centered on any ECAL/HCAL trigger tower pair.
In the Barrel, one ECAL trigger tower corresponds to an array of $5\times5$
crystals ($\Delta\eta\times\Delta\phi\simeq0.087\times0.087$) and
it is divided into strips of 5 crystals with common $\eta$ and different
$\phi$ values. In the Endcaps, the ECAL trigger towers are composed
of several pseudo-strips with variable shape and may extend over more
than one supercrystal. For $\left|\eta\right|$<1.74, the granularity
of the Endcap ECAL towers is the same as in the Barrel and the boundaries
of each HCAL trigger tower follow the corresponding ECAL trigger tower.
In the Endcap region with $\left|\eta\right|$>1.74 one HCAL trigger
tower has twice the $\phi$ dimension \cite{TriggerTDR}. The energy
readout of one HCAL trigger tower is longitudinally segmented into
up to three elements, depending on the $\eta$ region.

\begin{figure}[H]
\begin{center}\includegraphics[%
  width=0.70\textwidth]{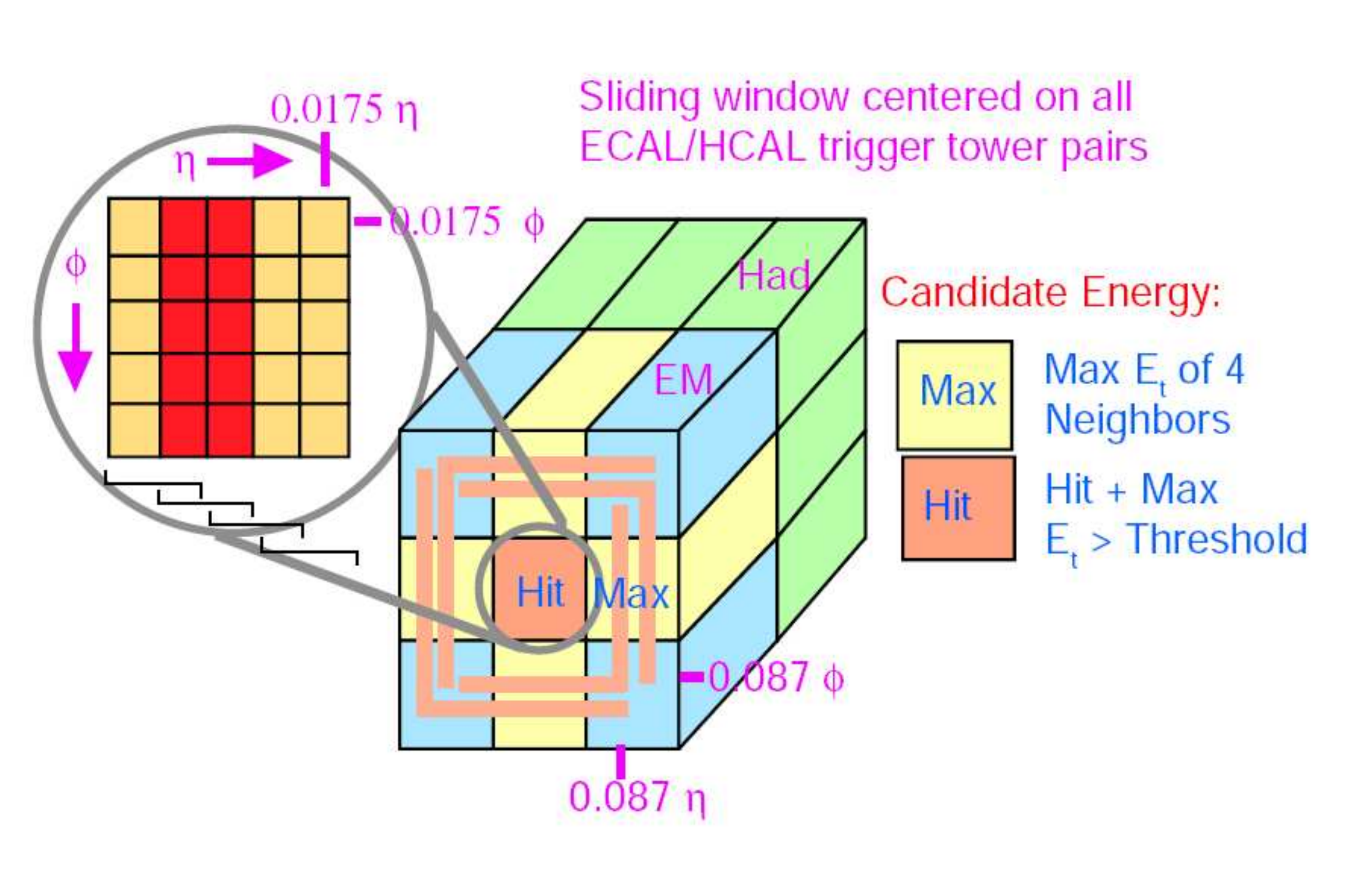}\end{center}

\caption{Sketch of Level 1 Electron/Photon trigger algorithm.\label{cap:figure1} }
\end{figure}

An illustration of the algorithm is shown in Figure \ref{cap:figure1}.
The algorithm estimates the candidate transverse energy $(E\mathrm{_{T}^{cand}})$
by summing the $E\mathrm{_{T}}$ in the central trigger tower (\emph{HIT}
Tower) with the maximum $E\mathrm{_{T}}$ of its four broad side neighbor
towers.  Two independent streams of candidates are considered :
\emph{isolated} and \emph{non-isolated} electron/photons. An electromagnetic
object is included in the non-isolated stream if it is not vetoed
by the following two discriminators:

\begin{itemize}
\item Fine Grain (FG) veto : The lateral extension of the energy deposit
of an electromagnetic shower in the ECAL is typically narrow, since
the Moliere radius and the crystal transverse dimensions are similar.
In the CMS experimental setup, due to the effects of magnetic field,
bremsstrahlung and photon conversion in the tracker material, the
shower spreads along $\eta$ and $\phi$ in the ECAL. An efficient
collection of all the particle energy requires a summation of the
crystal energy deposits along the $\phi$ bending direction of the
magnetic field. One must also take into account a smaller spread along
the $\eta$ direction, due to the intrinsic shower transverse development
and the ECAL entry point. The evaluation of the FG veto proceeds in
the following steps. First, the energy released in each pair of adjacent
strips is computed for each trigger tower. Then, the strip pair with
the maximum energy is found. The total energy released in the trigger
tower $(E\mathrm{\mathrm{_{ECAL}}})$ is also computed. Next, the
ratio of the strip pair with maximum energy to the total energy, the
Fine Grain ratio $(R\mathrm{^{FG}})$, is evaluated. Finally, $R\mathrm{^{FG}}$
is compared to a threshold $(R\mathrm{_{thr}^{FG}})$ and $E\mathrm{_{ECAL}}$
is compared to a minimum energy threshold $(E\mathrm{_{thr}^{FG})}$.
If $R\mathrm{^{FG}}<R\mathrm{_{thr}^{FG}}$ and $E\mathrm{_{ECAL}}>E\mathrm{_{thr}^{FG}}$,
a FG veto bit is set for the candidate. A typical value for $R\mathrm{_{thr}^{FG}}$
is $0.9$. The adequate value for $E\mathrm{_{thr}^{FG}}$ is discussed
in Section \ref{sec:Impact-of-Combined}. This definition of the algorithm
applies to the ECAL Barrel, whereas the Endcap different geometry
implies a different algorithm.
\item Hadronic Calorimeter (HAC) veto : The longitudinal profile of the
electron shower is measured by comparing the energy deposited in the
ECAL tower to the energy deposited in the corresponding HCAL tower
($E\mathrm{_{HCAL}})$. Indeed, the HAC veto bit is set if the energy
ratio $E\mathrm{_{ECAL}}/E\mathrm{_{HCAL}}$ $(R\mathrm{^{HAC}})$
is larger than a given threshold, $R\mathrm{_{thr}^{HAC}}$, and if
there is a significant activity in the ECAL trigger tower, that is,
if $E\mathrm{_{ECAL}}>E\mathrm{_{thr}^{HAC}}$. This requirement
suppresses background from charged pions that deposit a non negligible
amount of energy in the ECAL. Typical values for $R\mathrm{_{thr}^{HAC}}$
and $E\mathrm{_{thr}^{HAC}}$ are 0.05 and 3 GeV, respectively. If
$E\mathrm{_{ECAL}}<E\mathrm{_{thr}^{HAC}}$, the HAC veto is not applied.
Thus minimum ionizing pions not interacting in the ECAL are not rejected
by this veto alone.
\end{itemize}
The \emph{non-isolated} electron/photon stream uses the output of
the two previous sub-algorithms evaluated only for the \emph{HIT}
tower. The \emph{isolated} stream requires additional information
from the eight nearest neighbors around the \emph{HIT} trigger tower.
A candidate is included in the \emph{isolated} stream if it is not
rejected by the previous two vetoes and, furthermore, if it is not
vetoed by the following two conditions:

\begin{itemize}
\item Combined Fine Grain and HAC veto isolation (\emph{Neighbor veto})
: For each nearest neighbor tower of the \emph{HIT} tower the FG veto
and HAC veto bits, as defined earlier, are computed to yield a combined
veto bit. In the current implementation of the algorithm, the output
of the combined veto is a simple logical OR of the two vetoes. The
candidate is not included in the \emph{isolated} stream if any FG
or HAC veto bit of the eight nearest neighbor towers is set. That
is, a candidate is considered \emph{isolated} if it passes both the
FG and HAC veto on all eight nearest neighbors. 

\item ECAL isolation veto : From the eight nearest neighbor towers of the
\emph{HIT} tower, all possible sets of five contiguous towers with
three of them placed in the corners of the $3\times3$ window are
formed. There are four such sets, represented in Figure \ref{cap:figure1}
by orange {}``L-shaped'' lines. At least for one of these sets,
all towers must have energy below a given threshold $(E\mathrm{_{thr}^{iso}})$
for the candidate to be accepted. The isolation criterion is required
only in one corner to prevent self-veto of the electron candidates
due to a possible leakage of energy to the nearest neighbor tower.
Typical values of $E\mathrm{_{thr}^{iso}}$ are well above the trigger
tower noise level, e.g, 1.5 GeV. 
\end{itemize}

\subsection{Summary of L1 Hardware Implementation and Dataflow}

A detailed description of the hardware implementation and dataflow
of the L1 electron/photon algorithm can be found in Refs. \cite{CMS_Note_1999-026}
and \cite{TriggerTDR}. Here only a brief outline is presented.

The FG bit is evaluated for each trigger tower in the frond-end electronics
by the Trigger Primitive Generator (TPG) sub-system, which is implemented
in a FENIX ASIC chip. Both $R\mathrm{^{FG}}$ and $E\mathrm{_{ECAL}}$
are compared to two thresholds, which can be set to different values
for each run. The FG bit is then calculated using a lookup table (LUT)
whose inputs are the results of the previous comparisons. This setup
provides limited flexibility to make the FG bit dependent on $E\mathrm{_{ECAL}}$.
Illustration of part of the code is presented in Appendix \ref{sec:appendix A}.
Due to hardware design, it is not possible to set different thresholds
for the individual crystals in order to suppress very low energy deposits
comparable to noise level. The hardware implementation of the FG veto
employs strip and tower transverse energy instead of energy; however
the data used in this study was collected in the region of $\eta\sim0.2$,
where $\sin\theta\sim0.98$, so $E_{T}/E$ is close to unity and the
residual difference is further decreased by the ratio. 

The HAC veto, the Neighbor veto, and the ECAL isolation portions of
the algorithm are implemented in the Regional Calorimeter Trigger
(RCT). Two VME cards are used, the Receiver and the Electron ID card.
The Receiver card receives $E{\rm {_{C}}}$ and $H{\rm {_{C}}}$ ,
which are the ECAL and HCAL trigger tower energy sent via serial links
from the calorimeter TPGs on an eight-bit, compressed, non-linear
scale. Additionally, the FG bit from the ECAL TPG is also received.
These quantities are fed into a LUT whose output consists of seven
bits of transverse energy (e.g. $E_{{\rm ECAL}}+E_{{\rm HCAL}}$)
and one bit, nominally known as the electron veto bit, which is a
convolution of the HAC and FG veto bits. Again, all these bits are
evaluated separately for each trigger tower. The Electron ID card
receives these bits from all towers in each $4\times4$ trigger tower
region (calorimeter region) and : 1) estimates the candidate transverse
energy, 2) makes the OR of $e_{{\rm veto}}$ of the eight nearest
neighbors, and 3) evaluates the ECAL isolation veto bit. It can find
two possible electron candidates in each calorimeter region, corresponding
to two separate streams, with the nominal designation of \emph{isolated}
or \emph{non-isolated} .

The two streams mentioned earlier are exclusive. Trigger cuts are
not applied to the electron/photon candidates that saturate the L1
linear $E\mathrm{_{T}}$ scale ( \textasciitilde{} 63.5 GeV), which
are treated as if they were isolated. The thresholds of the electron
algorithm are set for all towers at the same values by default, but
can be programmed differently.

In each calorimeter region, the highest $E\mathrm{_{T}}$ electron/photon
candidates, either non-isolated or isolated, are separately found.
The eight candidates of each stream found in a regional trigger crate
are further sorted by transverse energy. The four highest $E\mathrm{_{T}}$
candidates of each category from each crate are then transferred to
the Global Calorimeter Trigger where only the top four candidates
are retained for processing by the CMS global trigger.

\section{Combined ECAL/HCAL test beam in the H2 experimental area \label{sec:Experimental-Setup}}

\subsection{Detector Configuration and Beam Conditions}

In 2006, a fully instrumented ECAL supermodule (SM09) was assembled
together with one HCAL Barrel wedge and three HO rings in the H2
test beam line area at CERN, forming one entire $\phi$ slice of the
calorimeter, in a configuration similar to the final CMS experimental
setup. The main difference with respect to the operation in the CMS
experiment is the absence of the tracker material in front of the
ECAL crystals and the absence of the external magnetic field. The
detector was exposed to beams of charged hadrons and electrons originating
from the collision of 400 GeV/c protons with production targets. The
detectors were mounted on a moving table to allow changing the relative
direction of the beam with respect to the detector. The beam line
operated in two energy ranges. In the very low energy mode, electrons
and charged hadrons were produced with momenta between 1 and 9 GeV/c.
In the high energy mode, electrons were produced with momenta between
15 and 100 GeV/c and charged hadrons with momenta between 15 and 350
GeV/c. The charged hadron beam contained an admixture of pions, kaons,
protons and antiprotons, whose composition varied with the beam momentum
\cite{CMS_Note_07-012}. At low momentum, there was also a significant
contamination of the hadron beam by electrons and muons. At the highest
momenta, contamination of the electron beam by pions was larger. The
studies were performed using electron samples with beam energies of
9, 15, 20, 30, 50 and 100 GeV and hadron samples with beam energies
of 3, 5, 7, 9, 15, 20, 30, 50 and 100 GeV.

\subsection{Electronic's Noise}

\begin{figure}[H]
\begin{center}\begin{tabular}{cc}
\includegraphics[%
  width=0.45\textwidth]{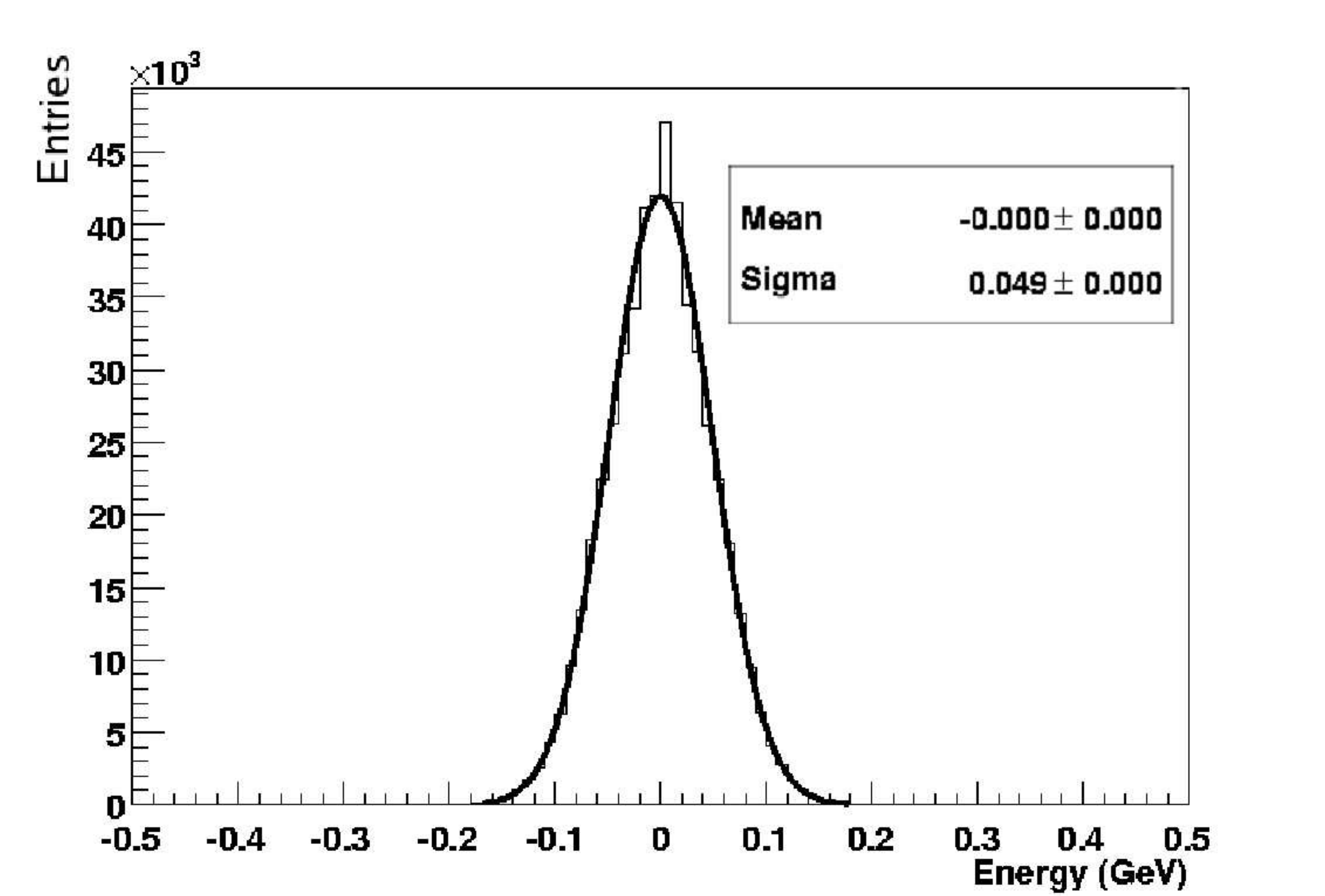}&
\includegraphics[%
  width=0.45\textwidth]{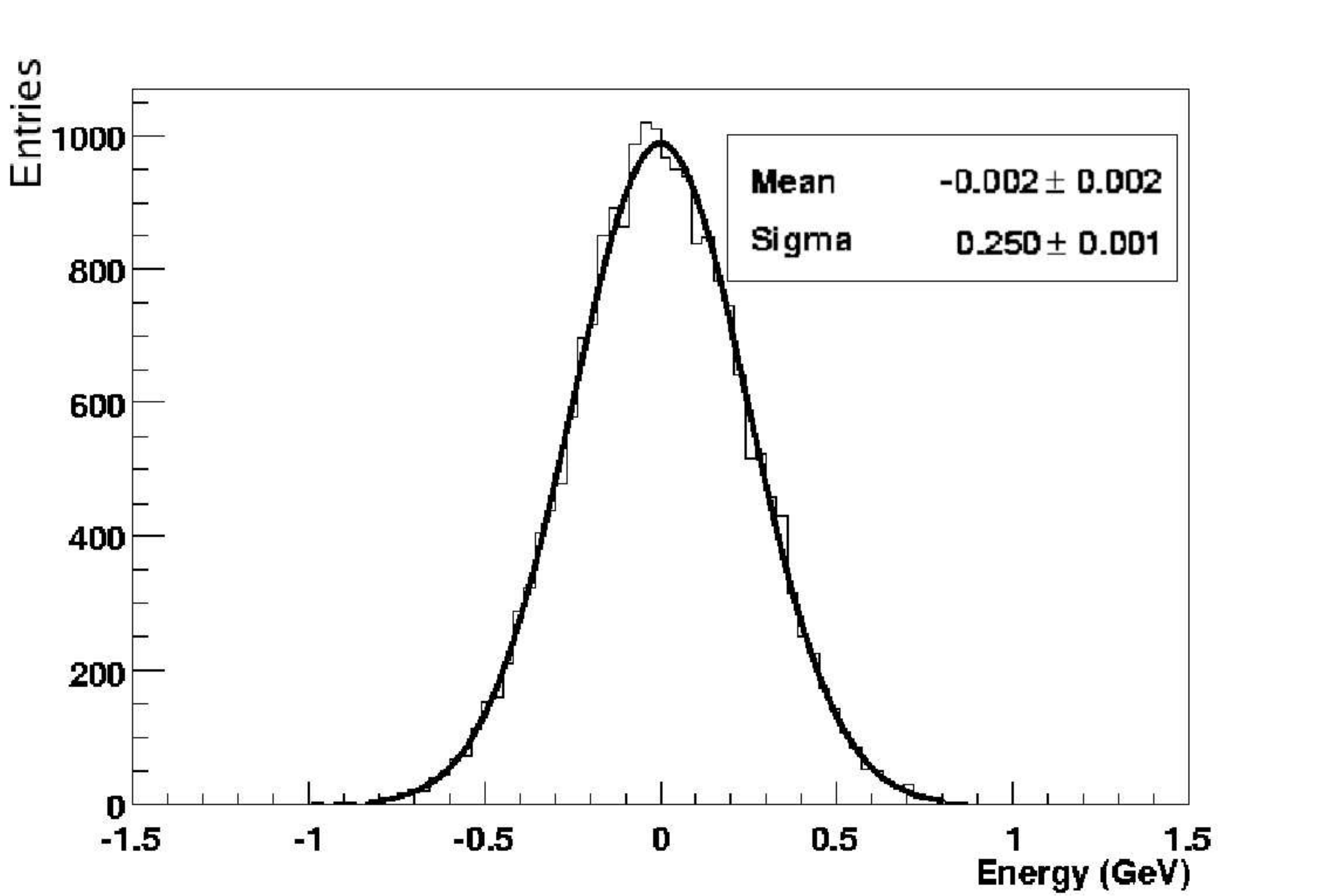}\tabularnewline
\end{tabular}\end{center}

\caption{Distributions of ECAL energy deposits in data taken while the beam
was not impinging on the detector for individual crystals (left)
and trigger towers (right) .\label{cap:Noise-level-for}}
\end{figure}

\begin{figure}[H]
\begin{center}\begin{tabular}{ccc}
a)\includegraphics[%
  width=0.32\textwidth]{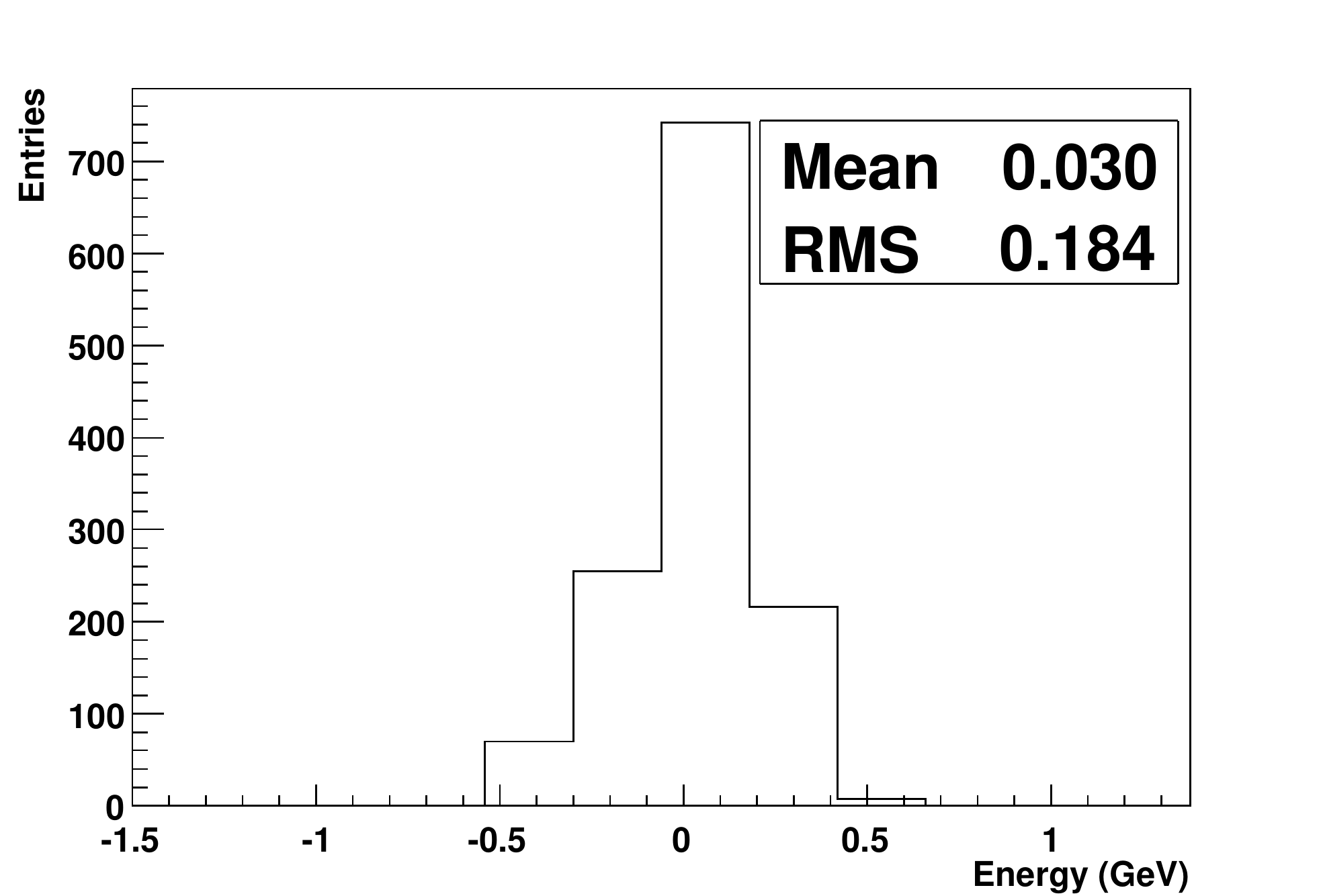}&
b)\includegraphics[%
  width=0.32\textwidth]{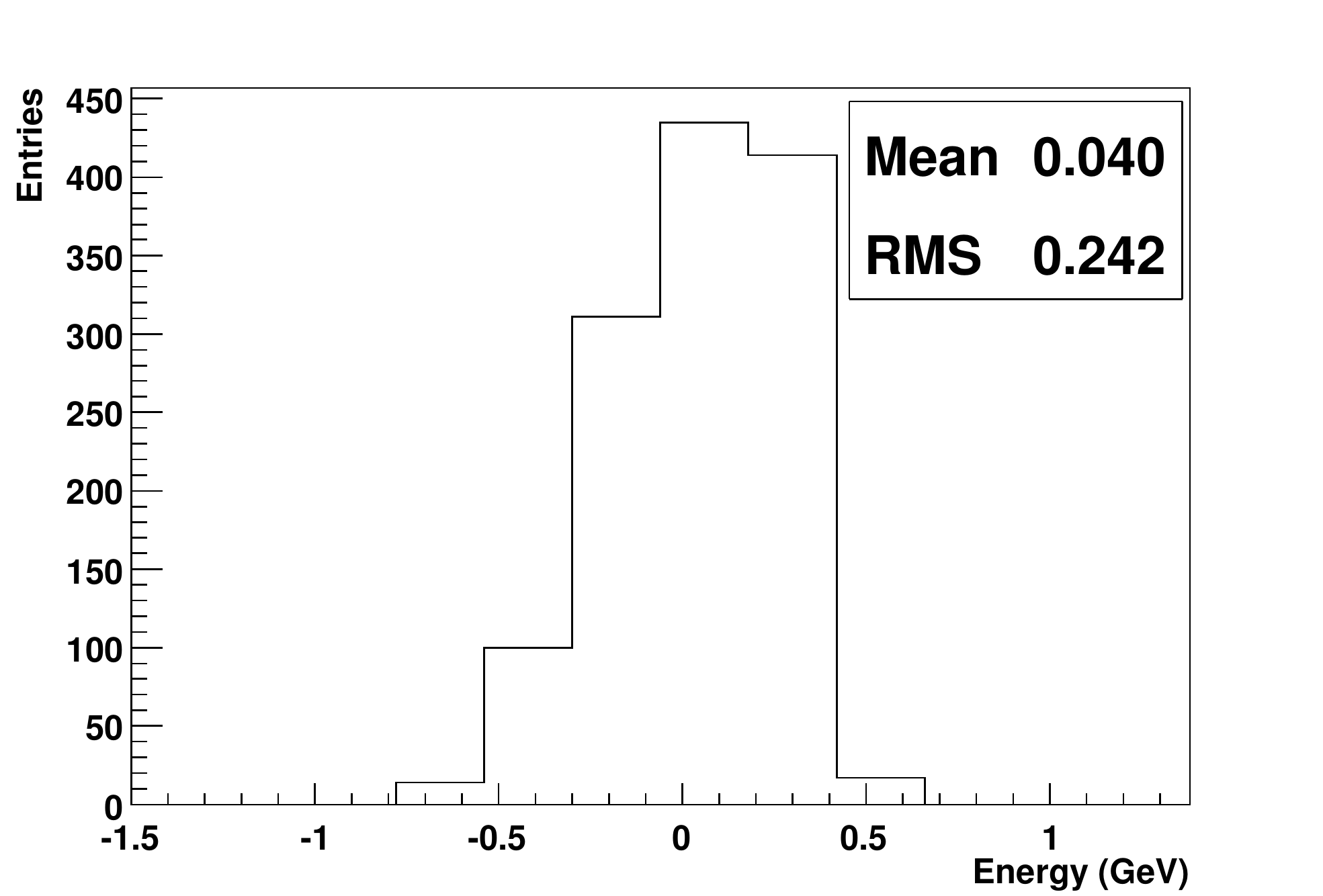}&
c)\includegraphics[%
  width=0.32\textwidth]{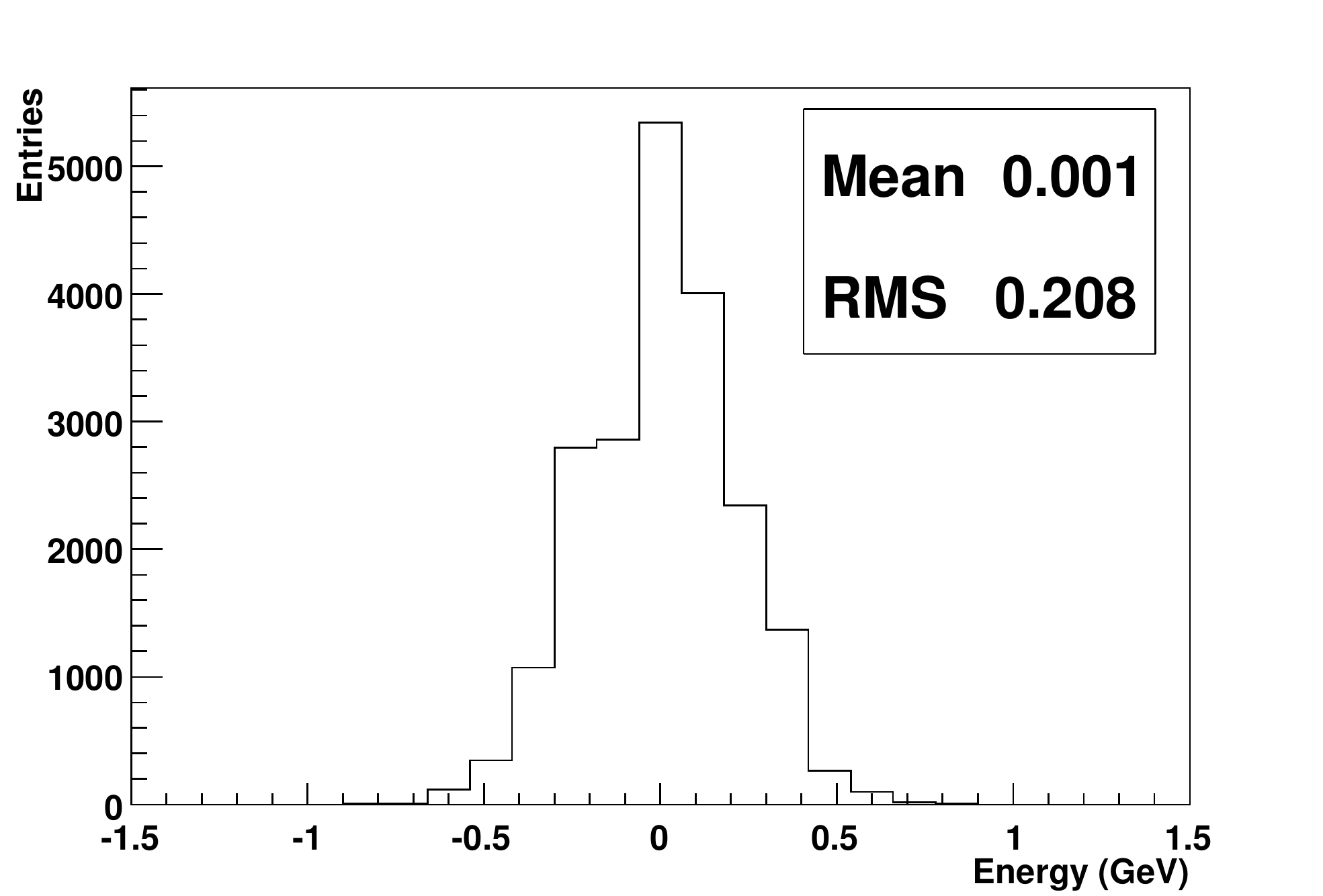}\tabularnewline
\end{tabular}\end{center}

\caption{Distributions of HCAL energy deposits in data taken while the beam
was not impinging on the detector for : trigger tower with ($\eta_{i}$,$\phi_{i}$)=3,13
(a), trigger tower with ($\eta_{i}$,$\phi_{i}$)=3,14 (b) and all
trigger towers of the first module (c).\label{cap:Distributions-of-energy}}
\end{figure}

The electronic's noise was estimated by measuring the energy deposits
of the individual calorimeter channels in data taken while the beam
was not impinging on the detector. The distribution of the average
energy deposit per crystal for the innermost ECAL module in $\eta$
of the supermodule ($20\times25$ crystal array in $\eta\times\phi$)
is shown in Figure \ref{cap:Noise-level-for} (left). The estimated
standard deviation of the fitted Gaussian distribution indicates a
$\sigma\mathrm{_{noise}}$\textasciitilde{} 50 MeV, close to the result
presented in an independent study \cite{CMS_Note_07-012} performed
in the same experimental conditions. Figure \ref{cap:Noise-level-for}
(right) shows the distribution of the average energy deposit per ECAL
trigger tower in the same region. The estimated $\sigma\mathrm{_{noise}}$
\textasciitilde{} 250 MeV is consistent with the result expected from
the noise sum of 25 independent crystal channels that constitute a
trigger tower. This value is different from the expectation of the
electronic's noise contribution for the ECAL TPG, due to the TPG linearization
and scale transformation of ECAL data and digitization effects. The
electronic noise of the HCAL readout energy is channel-dependent,
as can be seen in Figures \ref{cap:Distributions-of-energy}a and
\ref{cap:Distributions-of-energy}b, where the distribution of the
HCAL energy for two different trigger towers is shown. In Figure \ref{cap:Distributions-of-energy}c
the distribution of the HCAL energy per trigger tower averaged over
all the trigger towers of the first module is presented . The estimated
\emph{rms} is \textasciitilde{} 200 MeV.

\subsection{Calibration}

In this study the HCAL Barrel calibration constants obtained with
50 GeV/c electrons were used \cite{CMS_Note_07-012}. The calibration
was performed before the ECAL supermodule was mounted in front of
the HCAL detector, with the electron beam directed at the center of
each tower. Calibration constants of the ECAL supermodule obtained
from data collected with a beam of 50 GeV/c electrons pointing to
a grid of selected crystals were used. The crystal calibration constants
were calculated by minimizing the difference between the nominal electron
energy and the energy measured in a $5\times5$ crystal array (S25)
centered around the most energetic crystal, using a matrix inversion
technique \cite{PTDR1_4_43}. This method performs crystal inter-calibration
and sets the global energy scale at the same time. It is well suited
to \emph{in situ} calibration as it will be performed in the CMS experiment.
The signal amplitude measured in each crystal was reconstructed using
specific weights calculated for the ECAL supermodule SM09, according
to the method described in Ref.\cite{CMS_Note_06-037}. The calibration
samples and the data used in this study are independent. Despite the
non-compensating nature of the CMS calorimeter, no corrections to
the ECAL and HCAL calibrated energies were applied to make the combined
calorimeter response linear with energy.

\subsection{Data Selection}

\begin{figure}[H]
\begin{center}\includegraphics[%
  width=0.95\textwidth]{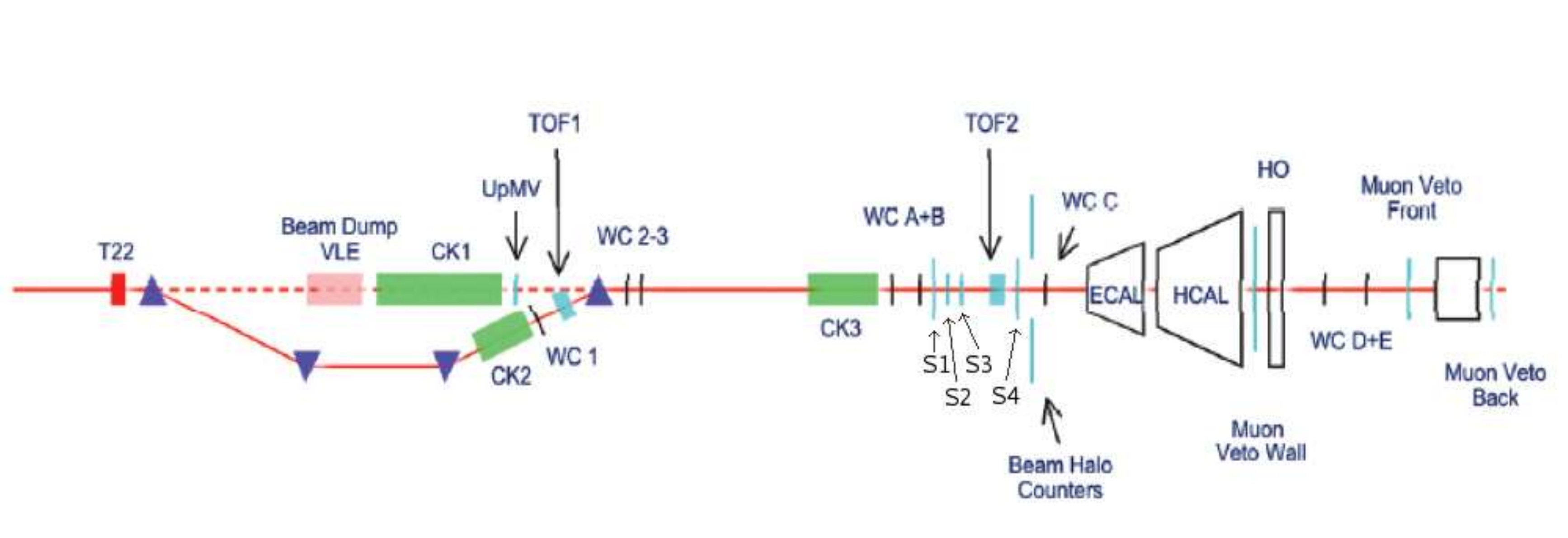}\end{center}

\caption{Schematic layout of the combined ECAL/HCAL test beam experimental
setup.\label{cap:The-Cern-H2}}
\end{figure}

A schematic layout of the H2 test beam experimental setup is shown
in Figure \ref{cap:The-Cern-H2}. In order to reduce spurious beam
contamination, the H2 test beam experimental setup included several
detectors dedicated to particle identification and beam cleaning,
placed along the beam line before the ECAL/HCAL detectors. Four scintillation
beam halo counters, arranged such that the beam passed through a $7\times7$
${\rm cm^{2}}$ opening, provided effective rejection of beam halo
and large-angle particles that originated from interactions along
the beam line. Additionally, requiring a single hit in the three
scintillation counters used as trigger (S1{*}S2{*}S4 coincidence)
allowed further rejection of multiple particle events. Energetic muons were tagged
by large scintillation counters placed well behind the calorimeters.
Soft muons were vetoed using an 80-cm thick iron block inserted in
front of the last muon counter and a muon veto wall, consisting of
8 individual scintillations counters, placed behind the HCAL Barrel.

In the low beam momentum mode, electrons and hadrons were discriminated
using the signal from CK2, a ${\rm \check{C}erenkov}$ threshold counter,
set to tag electrons only. At the lowest beam momenta ( $\lesssim$~3
GeV/c ), another ${\rm \check{C}erenkov}$ counter, CK3, provided
an electron double-tag. Although the signals from CK3 and two time-of-flight
counters could discriminate between pions, kaons and protons, the
information was not used in this study. In order to select electron
events, muon and hadron vetoes were required, whereas to select hadron
events, muon and electron vetoes were required, but no hadron discrimination
was applied.

\begin{figure}[H]
\begin{center}\begin{tabular}{cc}
\includegraphics[%
  width=0.49\textwidth]{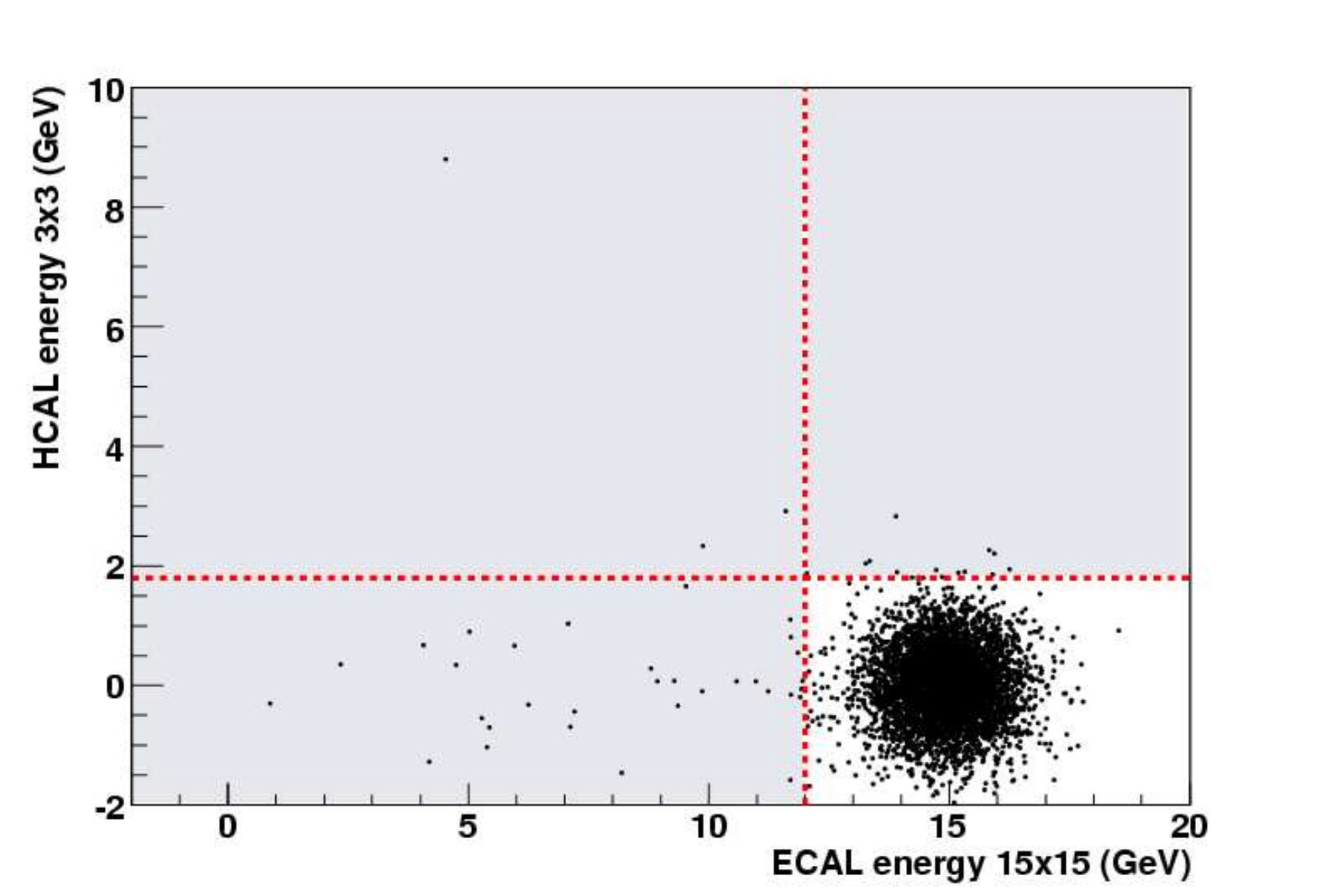}&
\includegraphics[%
  width=0.49\textwidth]{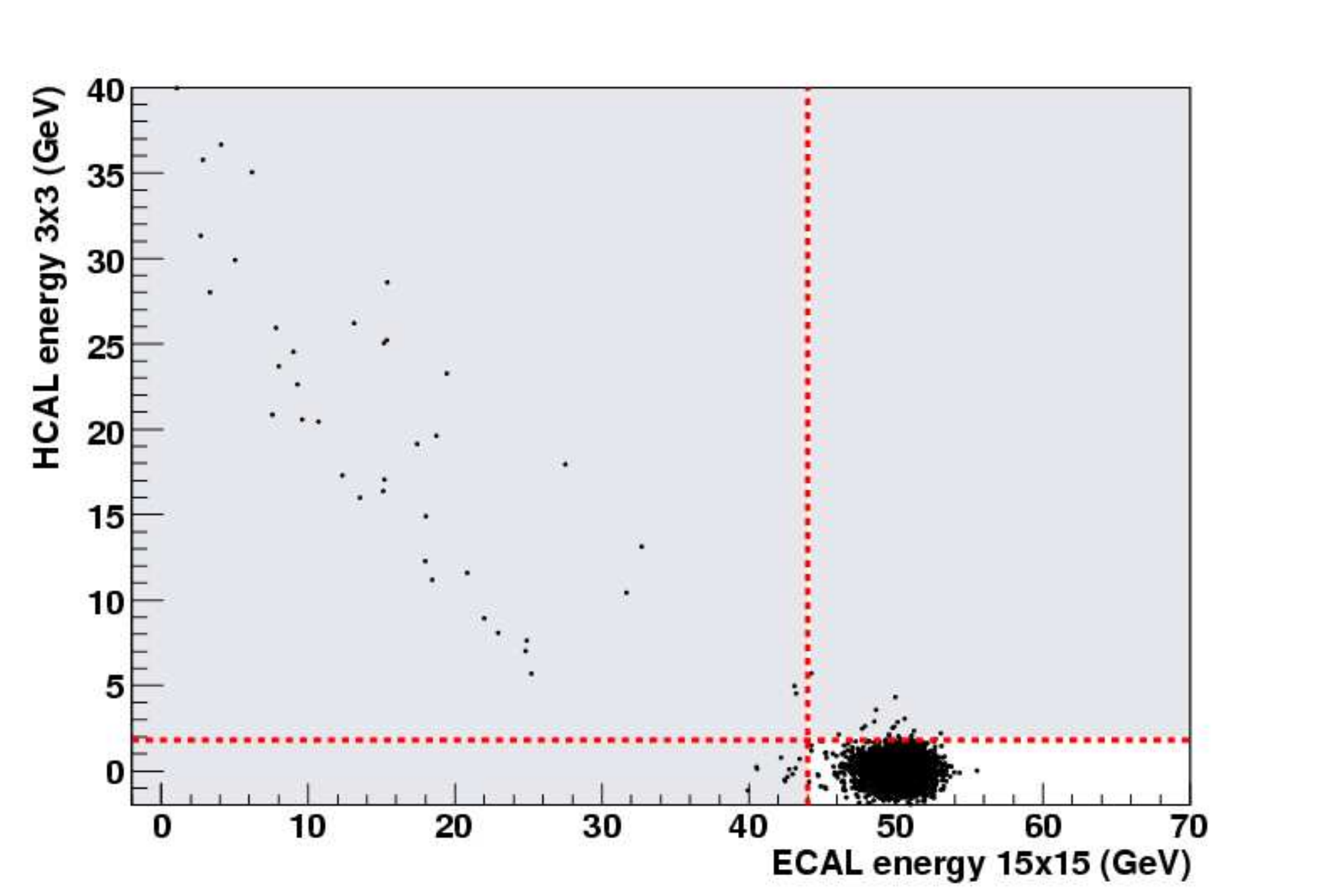}\tabularnewline
\end{tabular}\end{center}

\caption{Energy measured in $15\times15$ crystal array versus energy measured
in the corresponding $3\times3$ HCAL tower region for {}``electron''
beam events with 15 (left) and 50 (right) GeV nominal energy. Only
events with one reconstructed electromagnetic cluster are presented.
The dotted lines represent the selection cuts applied to obtain a
purified electron sample.\label{cap:selection1}}
\end{figure}

\begin{figure}[H]
\begin{center}\begin{tabular}{cc}
\includegraphics[%
  width=0.49\textwidth]{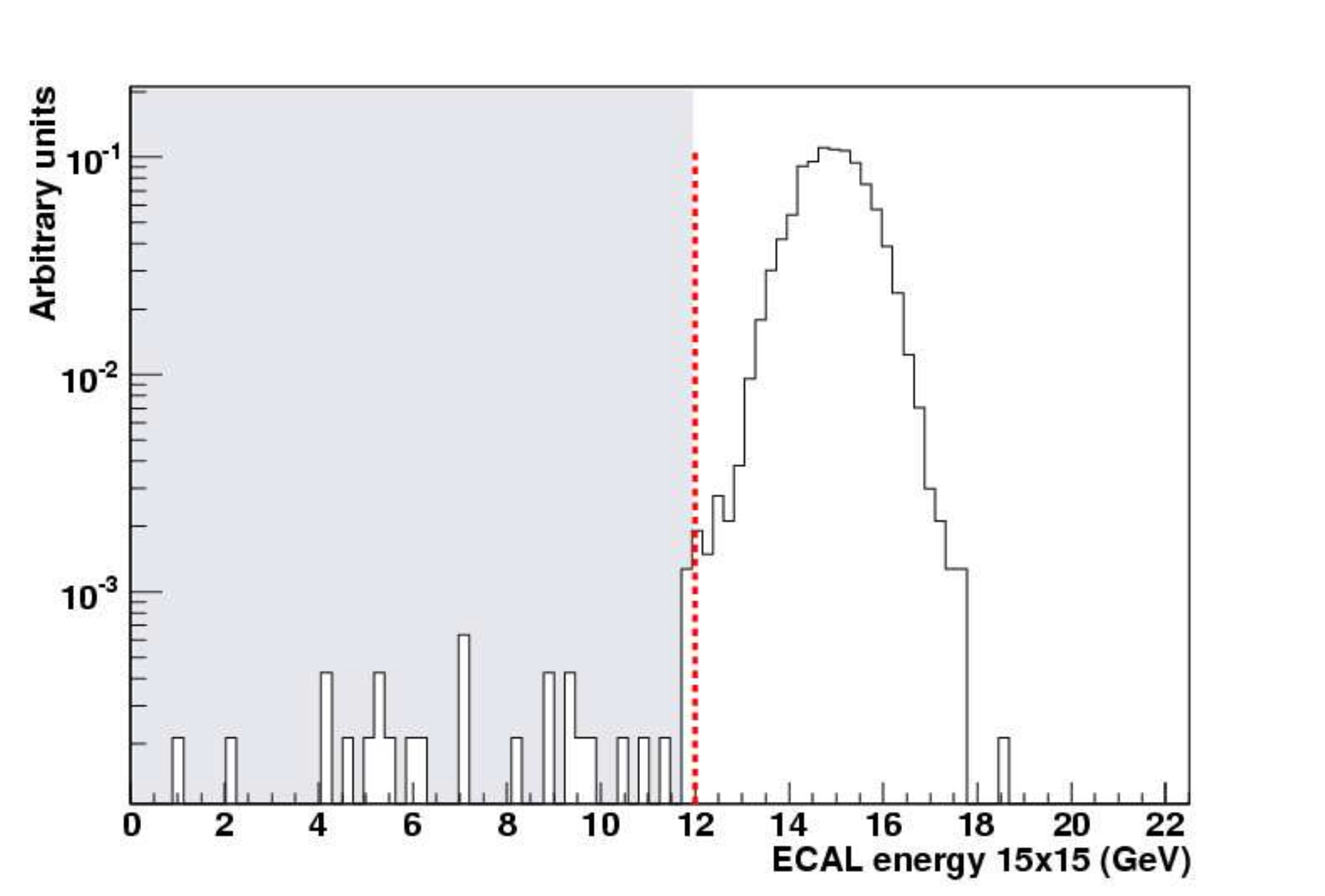}&
\includegraphics[%
  width=0.49\textwidth]{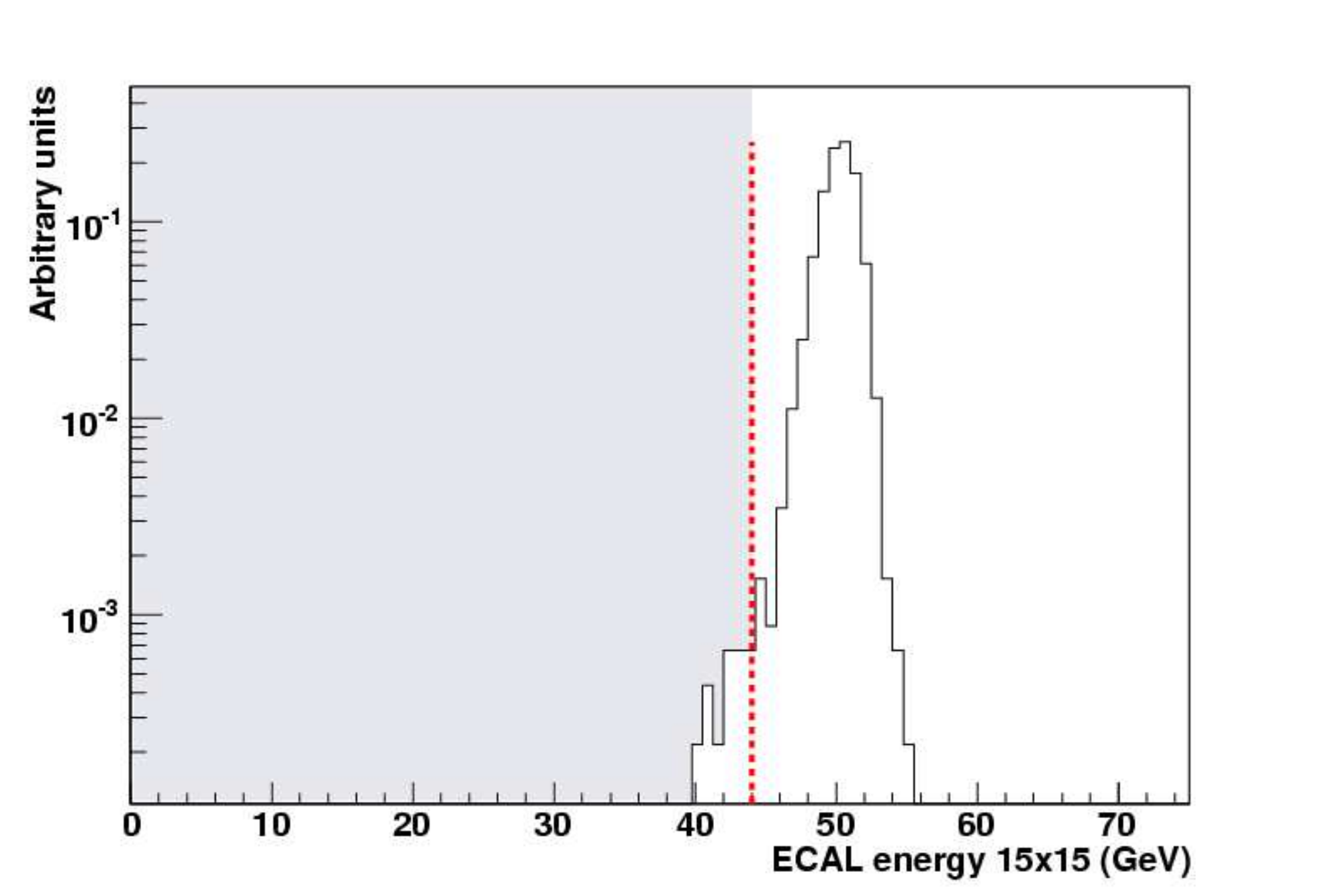}\tabularnewline
\end{tabular}\end{center}

\caption{Energy measured in $15\times15$ crystal array for {}``electron''
beam events with 15 (left) and 50 (right) GeV nominal energy. Only
events with measured HCAL energy in $3\times3$ tower region less
than 1.8 GeV and one reconstructed electromagnetic cluster are presented.
The dotted lines represent the selection cuts applied to obtain a
purified electron sample.\label{cap:selection2}}
\end{figure}

\begin{figure}[H]
\begin{center}\includegraphics[%
  width=0.60\textwidth]{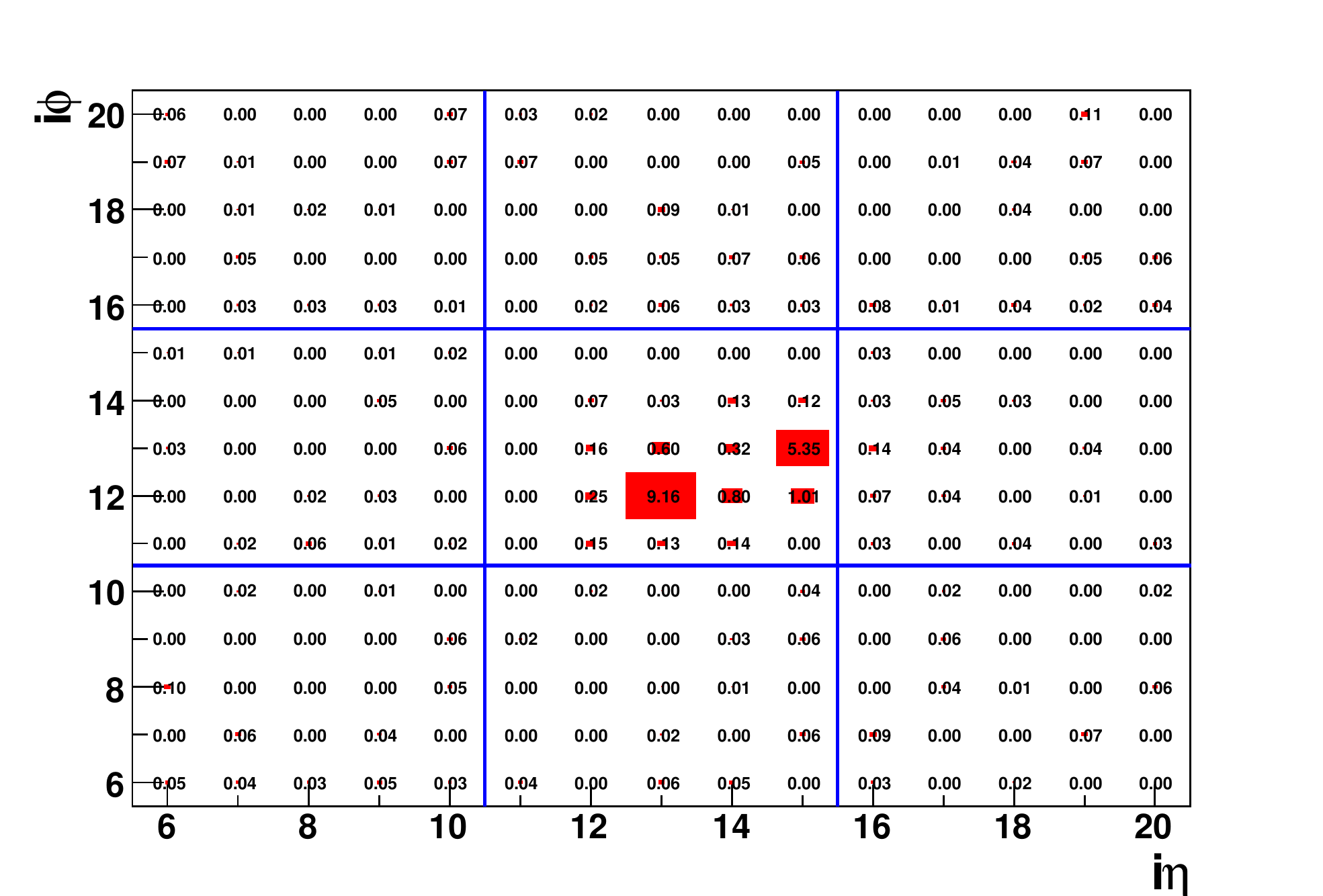}\end{center}

\caption{Individual ECAL crystal energy for a single event of electron beam
data sample with 20 GeV nominal energy. The two well- separated electromagnetic
clusters are most likely due to bremsstrahlung or a $\pi^{0}$ decays.\label{cap:Display-of-the}}
\end{figure}

In the high energy range, electron events were selected by applying
relaxed cuts on the ECAL energy versus HCAL energy plane as follows.
In order to reduce the bias due to energy containment effects, the
energy was measured in a large area of the ECAL detector, composed
of an array of $15\times15$ crystals, and in the corresponding $3\times3$
HCAL towers. There was no evidence of energy leakage of electrons
to the HCAL compartment for electrons with energies up to 100 GeV.
Therefore, events were required to have less than 1.8 GeV measured
in HCAL (equivalent to 3$\times$\emph{rms} of the HCAL noise distribution
for a $3\times3$ trigger tower region, assuming no noise correlation
between towers). Residual muon contamination was suppressed imposing
a lower limit on the measured ECAL energy of 2 GeV (the most probable
value of the distribution of the energy deposition of minimum ionizing
particles in the ECAL is \textasciitilde{} 300 MeV). Furthermore, a
higher but still loose lower limit was set on the measured ECAL energy,
with the intent of suppressing a small beam contamination from other
particles that interacted in the ECAL and left almost no energy in
HCAL. The value of this lower energy limit depends on the beam nominal
energy and is determined by the constraint to retain all the events
within a 3$\sigma\ $interval centered around the most probable value
of the energy distribution. The event selection based on ECAL/HCAL
energy cuts is illustrated in Figure \ref{cap:selection1} for 15
and 50 GeV beam energy. In Figure \ref{cap:selection2} it is shown
the ECAL energy measured in $15\times15$ crystal array imposing HCAL
$3\times3$ energy < 1.8 GeV. In some events two well separated electromagnetic
clusters could be found, possibly originating from electron bremsstrahlung
or $\pi^{0}$ decays. An example is shown in Figure \ref{cap:Display-of-the}.
The fraction of events with such topology depends on the data samples:
for one 9 GeV electron data sample it is \textasciitilde{} 5\%, while
for one 50 GeV electron data sample it amounts to 18\%. To systematically
reject those events, electromagnetic clusters were reconstructed using
the CMS Island algorithm \cite{CMS_Note_2001-034}, with an energy threshold for the seeds of
1 GeV. The Island algorithm makes electromagnetic clusters from
series of connected crystals containing energy deposits which decrease
monotonically starting from the highest energy and non adjacent seed
crystals. Therefore the number of clusters found correspond to the
number of local maxima in the array of crystal energy deposits.
Only events with one cluster were selected.

\section{Impact of energy leakage in electron identification\label{sec:Impact-of-Combined}}

\begin{figure}[H]
\begin{center}\begin{tabular}{cc}
\includegraphics[%
  width=0.49\textwidth]{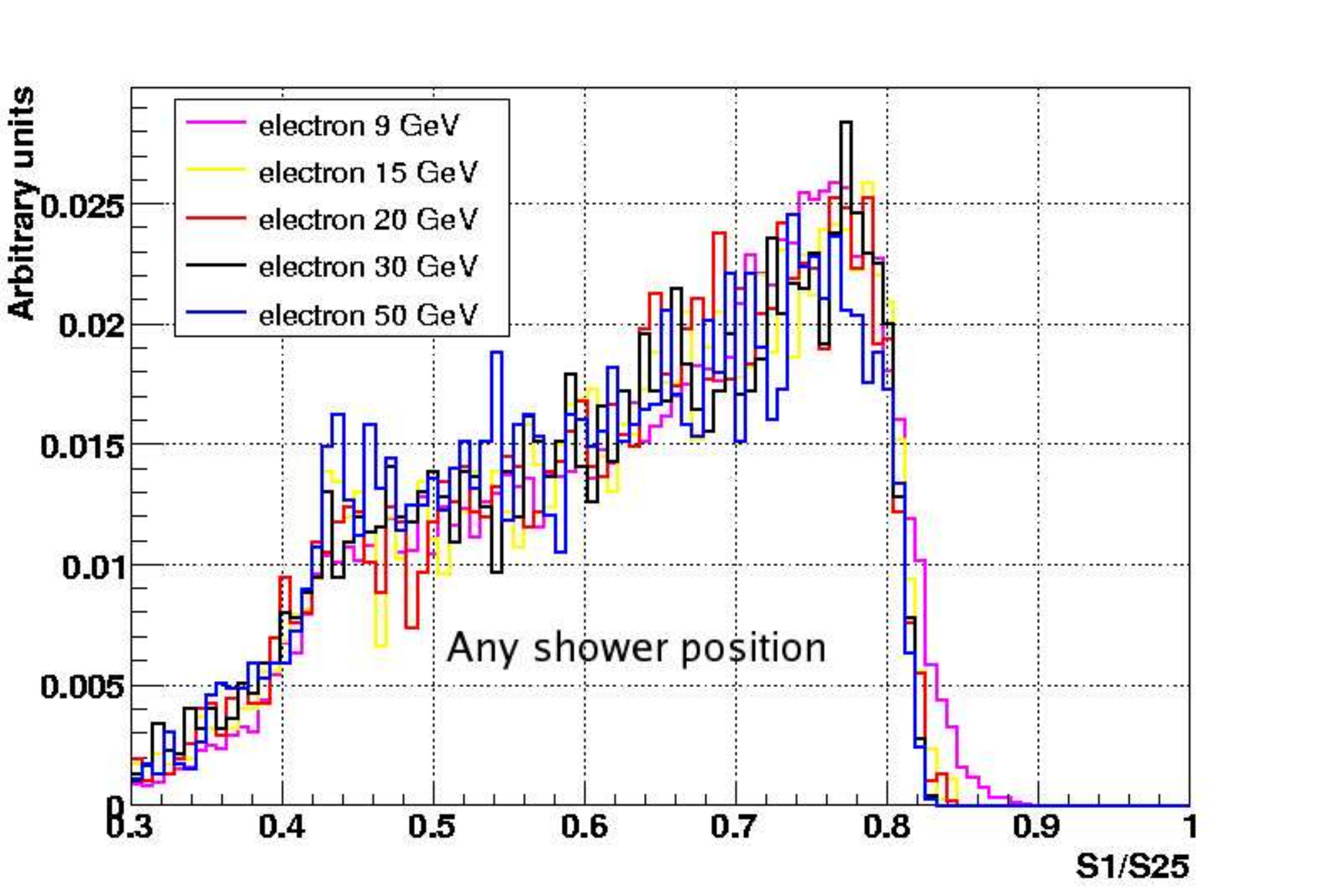}&
\includegraphics[%
  width=0.49\textwidth]{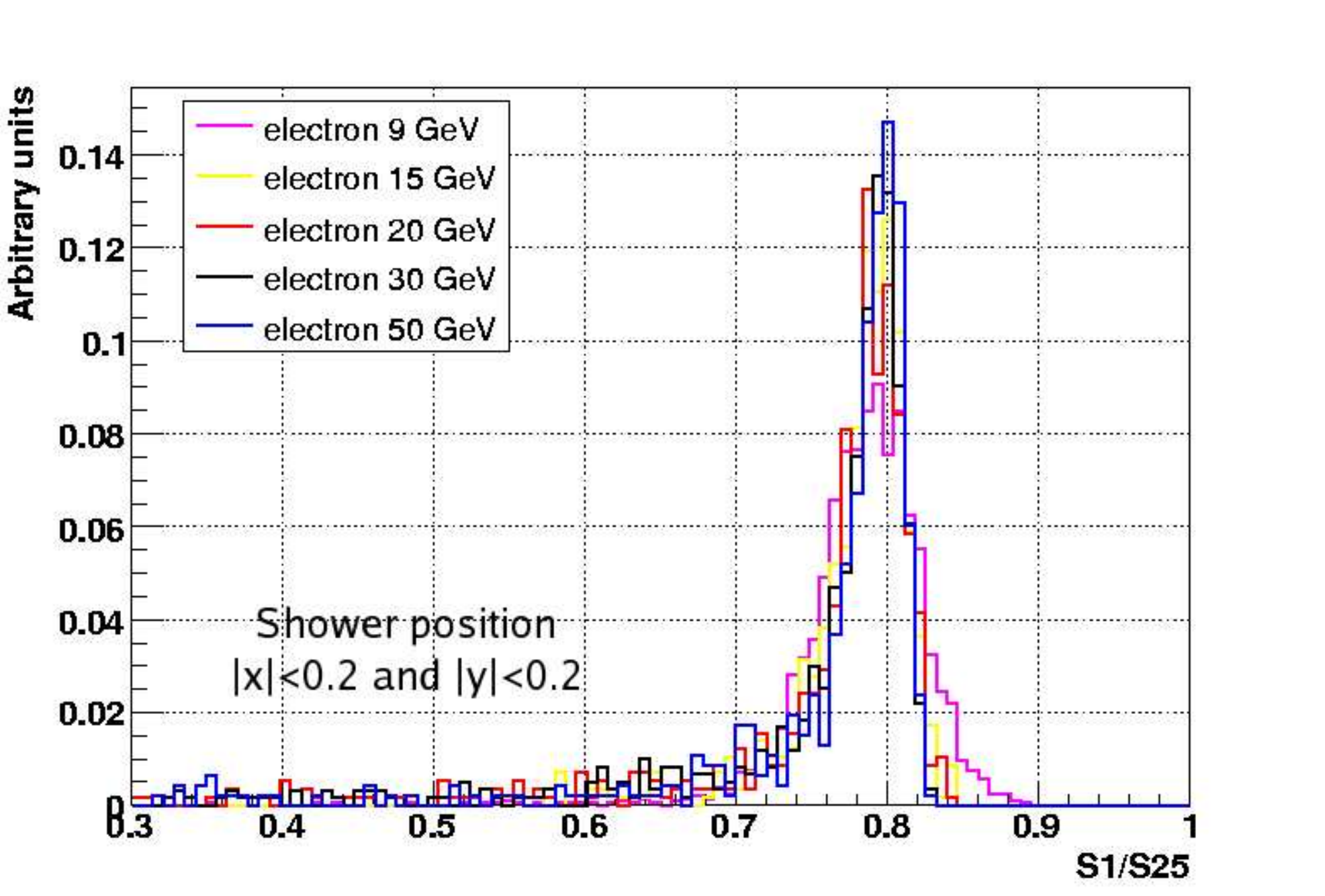}\tabularnewline
\includegraphics[%
  width=0.49\textwidth]{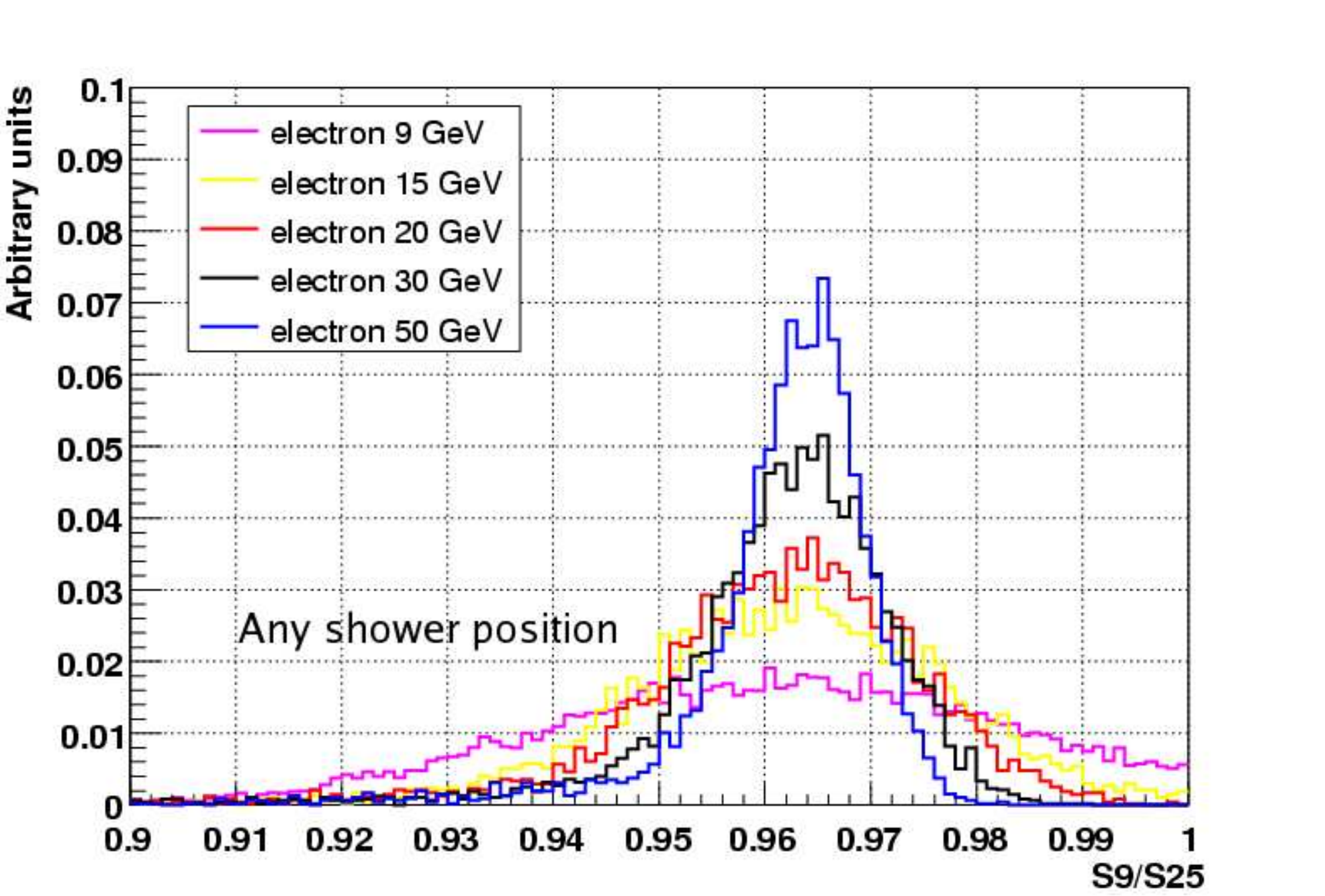}&
\includegraphics[%
  width=0.49\textwidth]{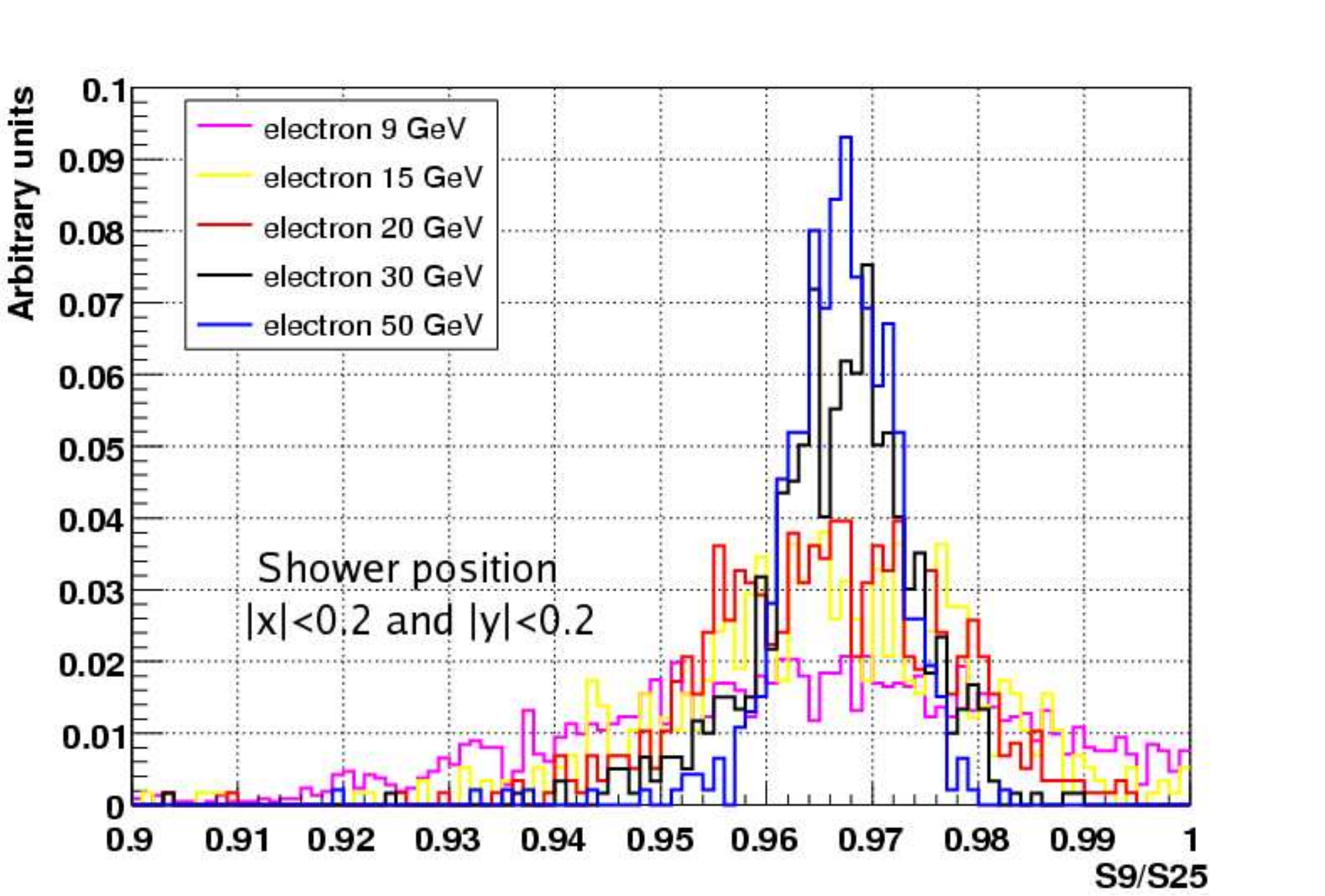}\tabularnewline
\end{tabular}\end{center}

\caption{Ratio of the energy measured in a single crystal (S1, top) and a
$3\times3$ crystal array (S9,bottom) to the energy measured in a
$5\times5$ crystal array (S25) for : all events (left), and events
with shower position $\left|x\right|<0.2$ and $\left|y\right|<0.2$
(right).\label{cap:Energy Fractions}}
\end{figure}

Electron showers deposit their energy in several crystals in the ECAL.
To reconstruct the electron energy one has to sum the energy deposited
in clusters of adjacent crystals. In test beam conditions, energy
sums of fixed arrays of crystals centered on the crystal having the
maximum energy give the best energy resolution \cite{PTDR1_4_32}.
Furthermore, the energy fraction contained in such clusters varies
with the shower position. The reconstructed energy fraction is maximum
when the shower is near the geometric center of the crystal, although
not exactly at the center due to the quasi-projective geometry of
the ECAL crystals. In Figure \ref{cap:Energy Fractions}
plots are shown for the ratio of the energy measured in a single crystal
(S1, top) and a $3\times3$ crystal array (S9, bottom) to the energy
measured in a $5\times5$ crystal array (S25), for different beam
energies. To estimate the shower position a weighted mean of the crystal's
position is used. The weights are given by the logarithm of the ratio
of the crystal energy to the total energy measured in a $3\times3$
crystal array centered on the most energetic crystal. We have chosen
a (X,Y) coordinate system, where X is parallel to $\eta$ and Y is
parallel to $\phi$; a crystal has a front-face size $\Delta x\times\Delta y=1\times1$
and the center of the most energetic crystal in each event is at $(x,y)=(0,0)$.
In Figure \ref{cap:Energy Fractions}, left panels refer to all events
and right panels refer only to events where the shower is closer to
center of the most energetic crystal, with $\left|x\right|<0.2$ and
$\left|y\right|<0.2$. In the top left panel one can see that the
S1/S25 energy fraction, regardless of the shower position, is lower
than \textasciitilde{} 80\% for all beam energies. Constraining the
shower position to the central region of the crystal, the lower tail
of the energy fraction distribution decreases significantly (Fig.
\ref{cap:Energy Fractions}, top right). The energy ratio S9/S25 is
closer to one, peaked at 96--97\% and its spread is much smaller
(Fig. \ref{cap:Energy Fractions}, bottom left). The decrease of the
spread of the energy fraction contained in the cluster as the beam
energy increases is correlated to the energy resolution of the ECAL,
which improves with energy.

The energy measured in a single ECAL trigger tower depends on the
impact position of the incident electron. In Figure \ref{cap:Candidate-Energy,-Energy}
(left) the energy response of a single trigger tower for events in
which the most energetic crystal is located inside (dotted line) or
outside (solid line) the inner $3\times3$ crystal array of the \emph{HIT}
tower (as defined in right panel of Figure \ref{cap:Fraction-of-events})
is shown for an electron beam sample with a nominal energy of 30 GeV.
An energy leakage is found when the shower maximum energy is deposited
near the boundary of the trigger tower. Furthermore, the energy resolution
for this class of events is also degraded. Adding the energy measured
in the \emph{HIT} tower to the energy measured in the second most
energetic tower (\emph{MAX} tower), the ECAL energy response becomes
similar for both classes of events and the energy resolution is improved,
as can be seen in Figure \ref{cap:Candidate-Energy,-Energy} (right).
The latter (\emph{HIT}+\emph{MAX} tower) is how the L1 electron trigger
algorithm calculates the electron candidate energy, as mentioned in
Section \ref{sub:Overview-of-Conceptual}.

\begin{figure}[H]
\begin{center}\begin{tabular}{cc}
\includegraphics[%
  width=0.49\textwidth]{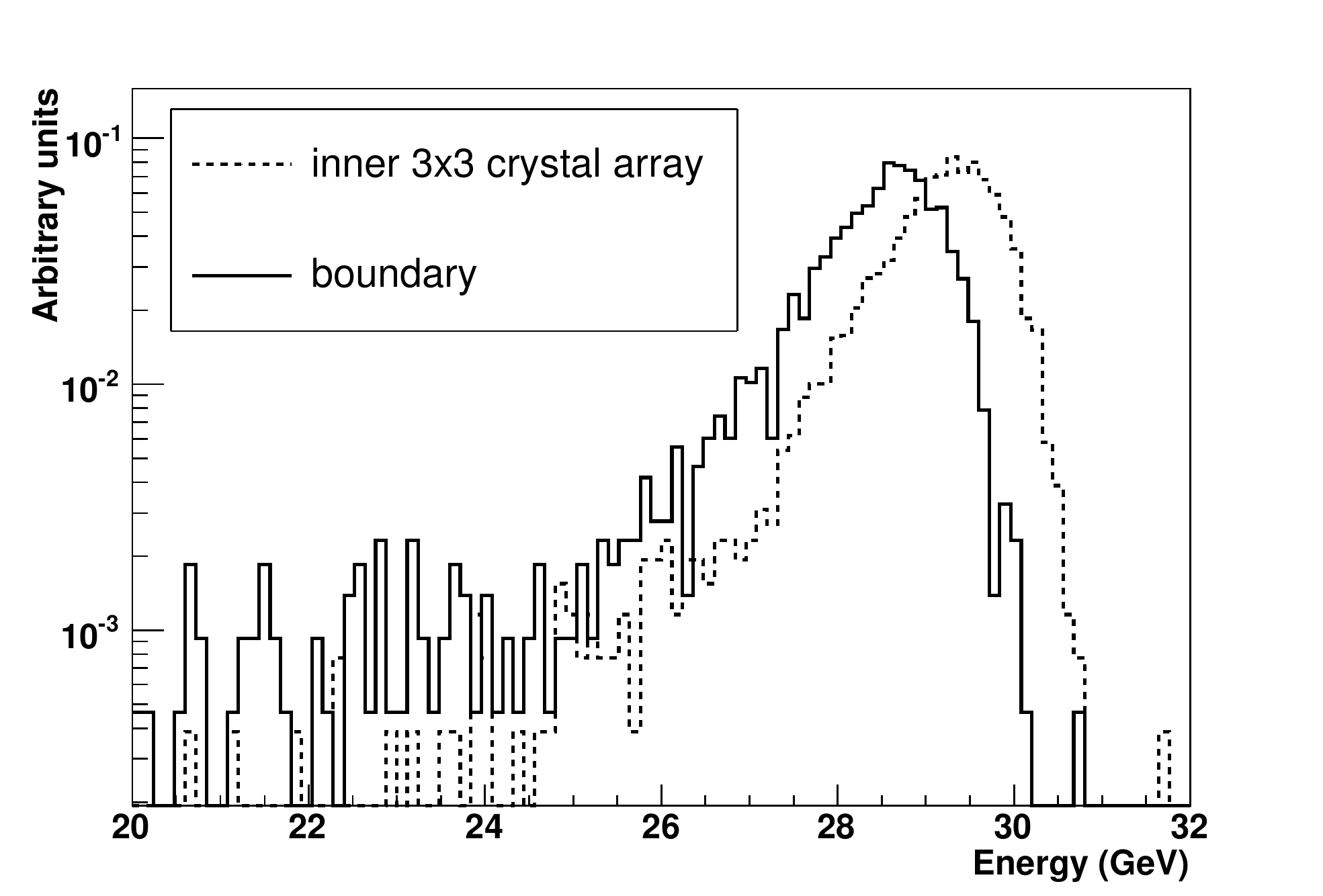}&
\includegraphics[%
  width=0.49\textwidth]{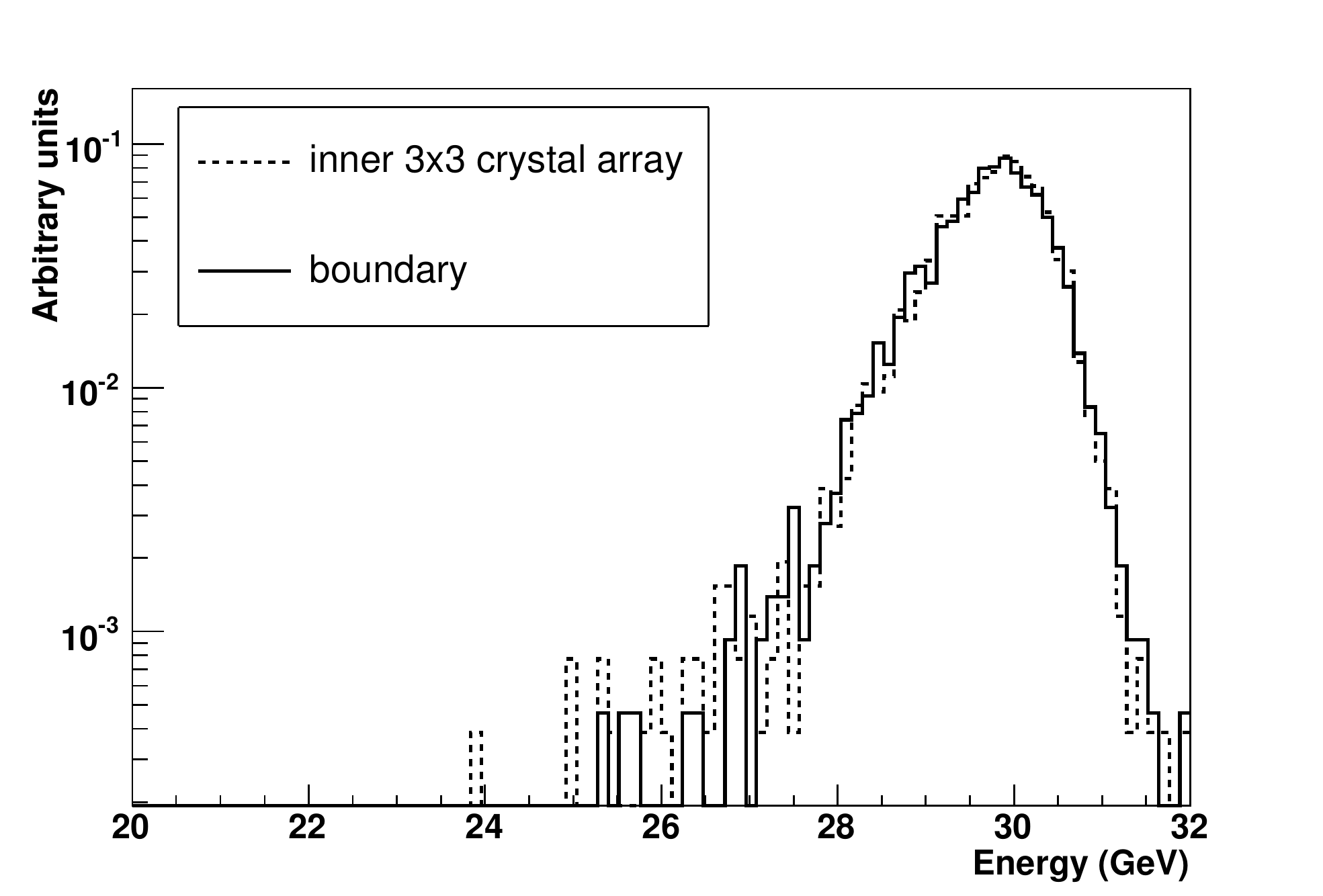}\tabularnewline
\end{tabular}\end{center}

\caption{Measured energy of a 30 GeV electron beam in: HIT tower (left) and
\emph{HIT} tower + \emph{MAX} tower (right) , for events in which
the most energetic crystal is located inside ( dotted line) or outside
(solid line) the inner $3\times3$ crystal array of the \emph{HIT}
trigger tower. \label{cap:Candidate-Energy,-Energy}}
\end{figure}

\begin{figure}[H]
\begin{center}\begin{tabular}{cc}
\includegraphics[%
  width=0.49\textwidth]{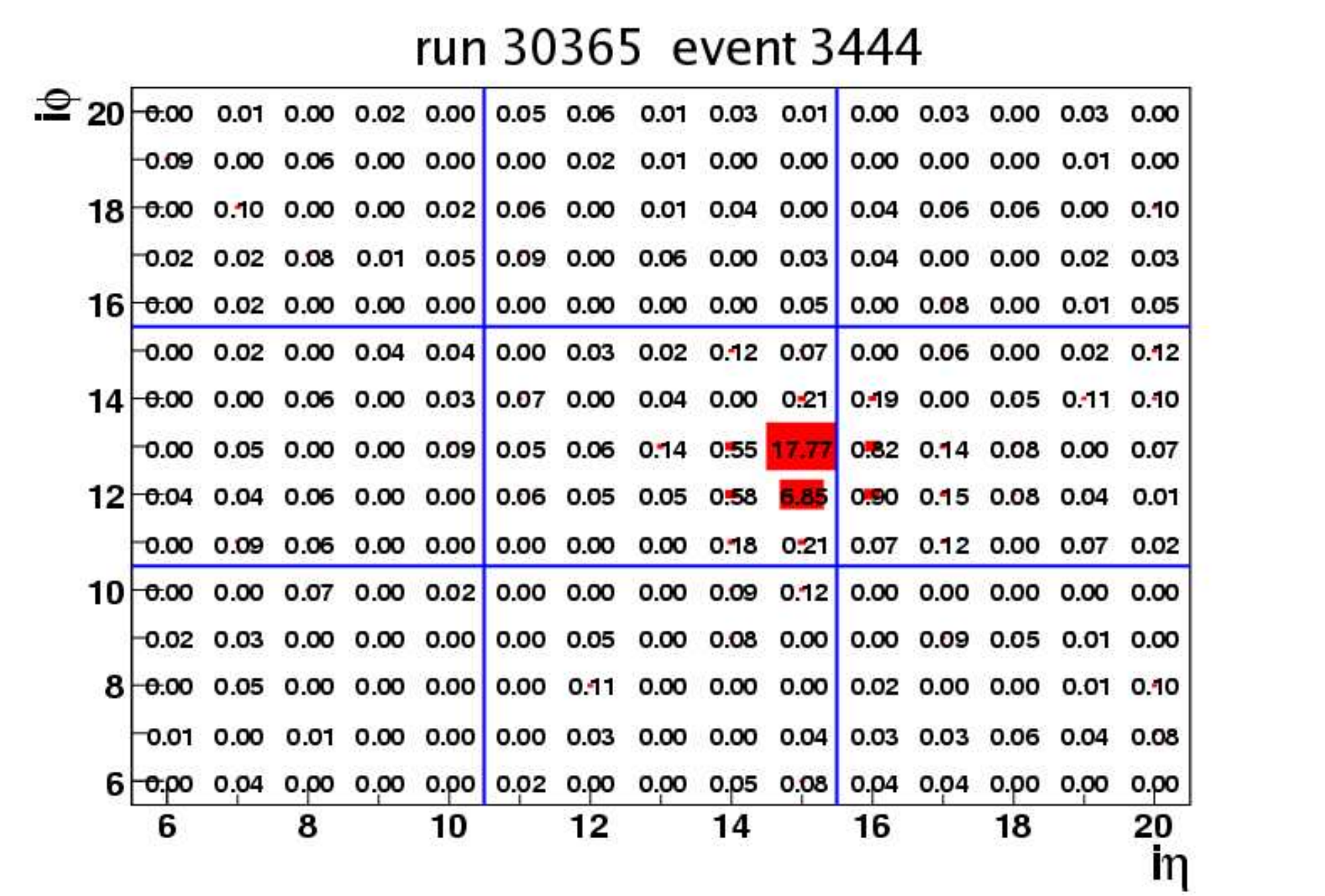}&
\includegraphics[%
  width=0.49\textwidth]{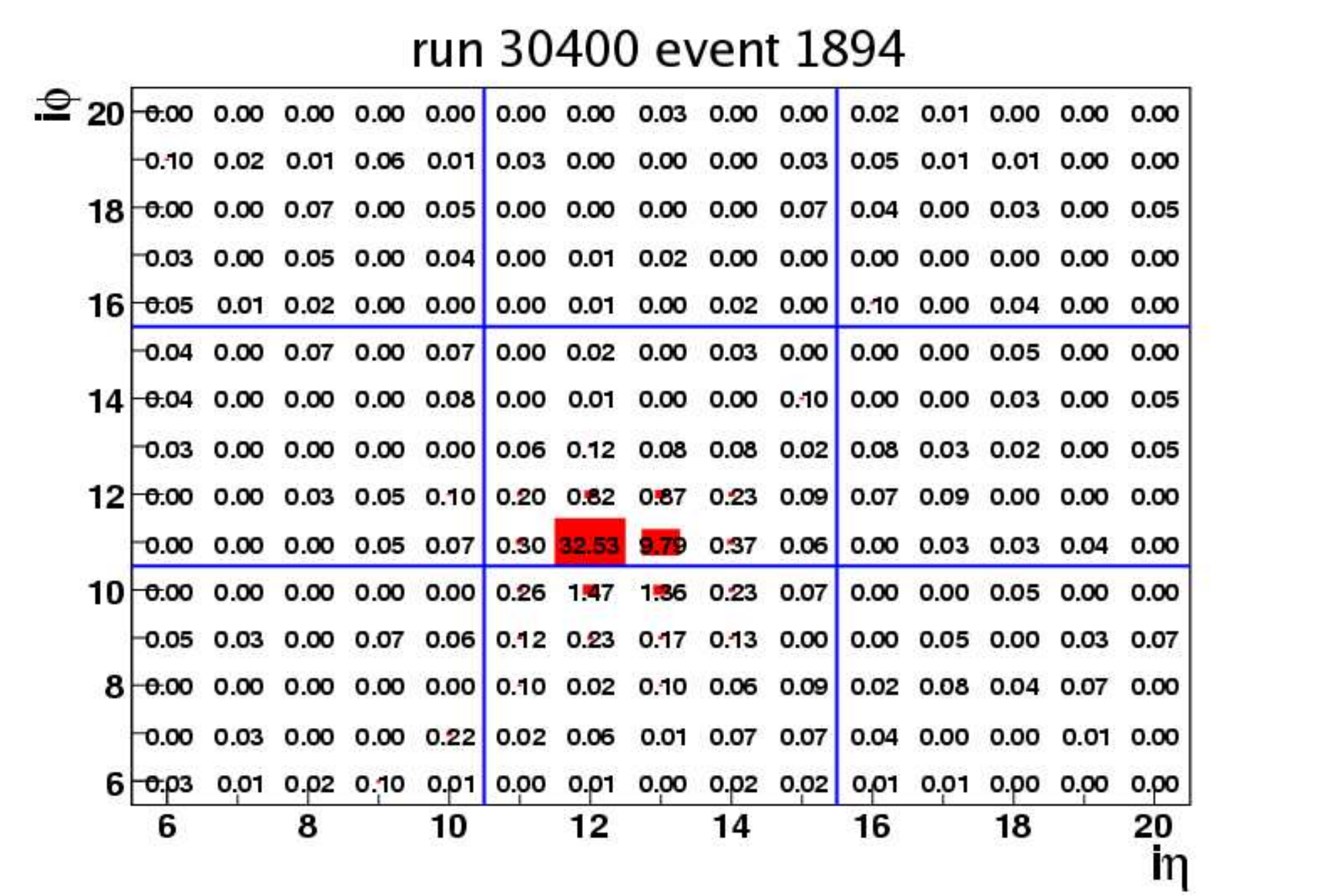}\tabularnewline
\end{tabular}\end{center}

\caption{ECAL crystal energy deposits for one event in: the electron beam
data sample with 30 GeV nominal energy (left) and the electron beam
data sample with 50 GeV nominal energy (right). The numbers in the
plots correspond to the energy (in GeV) measured in each crystal.\label{cap:figure2}}
\end{figure}

In the L1 electron/photon algorithm implementation, the FG veto is
evaluated individually for each trigger tower. Any energy leakage
outside of a trigger tower could lead to a trigger misclassification
of isolated electrons when calculating the OR of the FG bits of all
neighbor towers for the evaluation of the Combined Neighbor Veto bit.
Figure \ref{cap:figure2} shows two event displays of test beam data
where the energy leakage outside the \emph{HIT} trigger tower is manifest.
The solid lines delimit the trigger tower boundaries. Both events
correspond to electron beam data samples in which the cluster energy
deposits are compatible with an isolated candidate classification,
but fail the isolation criteria due to the energy leakage into the
neighboring trigger towers. Both events would be vetoed by the Combined
Neighbor Veto and considered as non-isolated, because the $R\mathrm{^{FG}}$
for the most energetic adjacent tower is below 0.8, as the threshold
is typically set at that value. The display on the right refers to
one event of an electron beam with nominal energy of 50 GeV. The tail
of the electromagnetic shower extends to the adjacent neighboring
tower leading to a spread of low energy deposits over more than 2
strips. The display on the left refers to one event of an electron
beam with nominal energy of 30 GeV. The low $R\mathrm{^{FG}}$ for
the neighbor tower might be caused by the conjunction of the low energetic
response of the shower leakage with noise fluctuation. These observations
show that care must be taken to ensure that spurious veto bits are
not set for low energy towers that may surround the \emph{HIT} tower,
either due to noise fluctuations, pile-up effects, energy leakage
outside the \emph{HIT} tower, or their association.

\begin{figure}[H]
\begin{center}\begin{tabular}{cc}
\includegraphics[%
  width=0.49\textwidth]{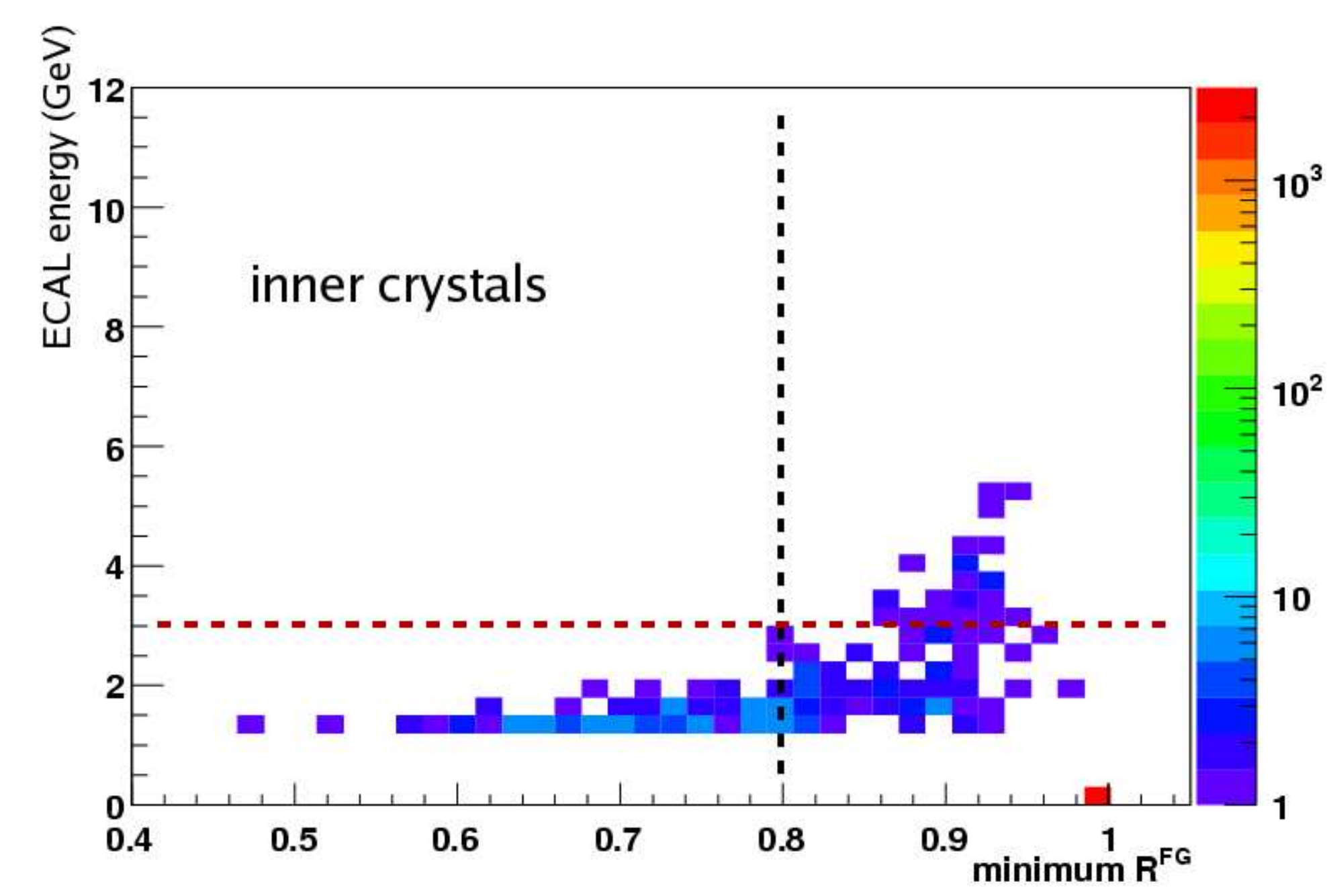}&
\includegraphics[%
  width=0.49\textwidth]{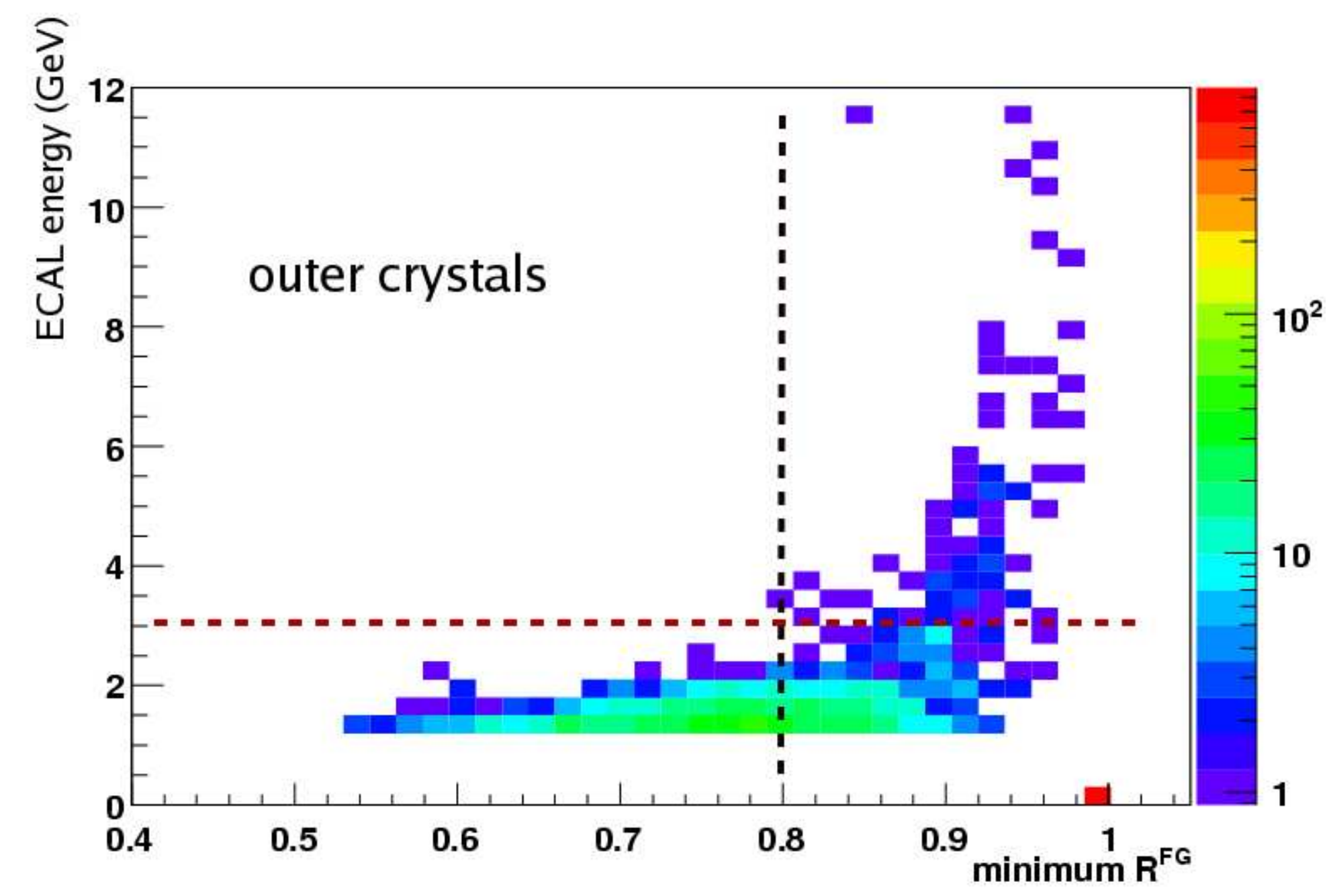}\tabularnewline
\end{tabular}\end{center}

\caption{FG ratio vs ECAL energy in the neighbor tower with minimum FG ratio
for events in which the most energetic crystal of the HIT tower is
located in the inner $3\times3$ crystal array (left), and on the
boundary crystals, outside the inner $3\times3$ crystal array (right).
\label{cap:figure4}}
\end{figure}

A correlation is found between the position of the most energetic
crystal within the \emph{HIT} trigger tower and the average minimum
value of $R\mathrm{^{FG}}$ for the neighbor towers. In Figure \ref{cap:figure4}
two scatter plots are shown for a data sample of 20 GeV electrons.
For each event the neighbor tower with minimum value for $R\mathrm{^{FG}}$
was determined. A threshold of 1.25 GeV (corresponding to a 5$\sigma$
noise level of trigger tower energy) was applied to the total energy
of the trigger towers, so that neighboring towers with lower energy
were not considered. The plots show the value of $R\mathrm{^{FG}}$
versus the ECAL energy for the neighboring tower. If no neighboring
towers with energy above 1.25 GeV are found, the $R\mathrm{^{FG}}$
is set to one and the ECAL tower energy is set to zero (i.e. no energy
above noise level, red bin in lower right part of each plot). The
left plot refers to events in which the most energetic crystal is
located in the inner $3\times3$ crystal array of the \emph{HIT} trigger
tower, whereas the right plot refers to events in which the most energetic
crystal belongs to the boundary of the \emph{HIT} tower, i.e., outside
the inner $3\times3$ crystal array (as defined in right panel of
Figure \ref{cap:Fraction-of-events}). The position of the most energetic
crystal is used as a rough estimator of the shower position with respect
to the \emph{HIT} trigger tower boundaries. It is seen that when the
maximum energy deposit is near the boundary of the \emph{HIT} tower
(Figure \ref{cap:figure4} right) more events have low $R\mathrm{^{FG}}$
measured value ($R\mathrm{^{FG}}\leq0.8$) for the neighbor tower
than when it is closer to the center of the trigger tower (Figure
\ref{cap:figure4} left). The estimated fraction of events is 27\%
and 3\%, respectively. The former observation is in agreement with
the hypothesis that low values of $R\mathrm{^{FG}}$ for the neighbor
towers are correlated to energy leakage effects. Furthermore, in both
cases all the events with low $R\mathrm{^{FG}}$ measured value deposit
relatively low energy in the neighboring tower, i.e. less than \textasciitilde{}
3 GeV for a 20 GeV electron beam.

\begin{figure}[H]
\begin{center}\begin{tabular}{cc}
\includegraphics[%
  width=0.49\textwidth]{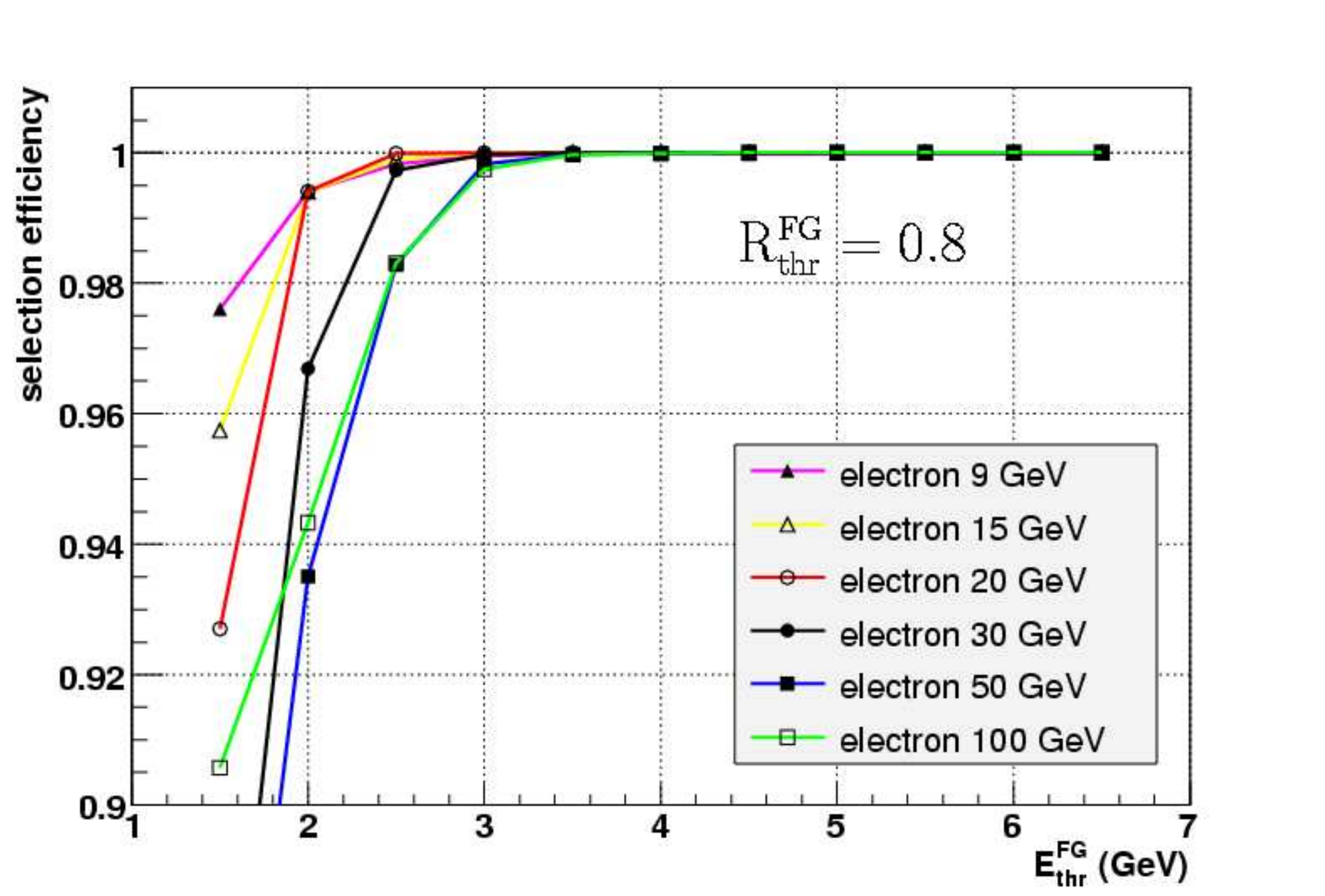}&
\includegraphics[%
  width=0.49\textwidth]{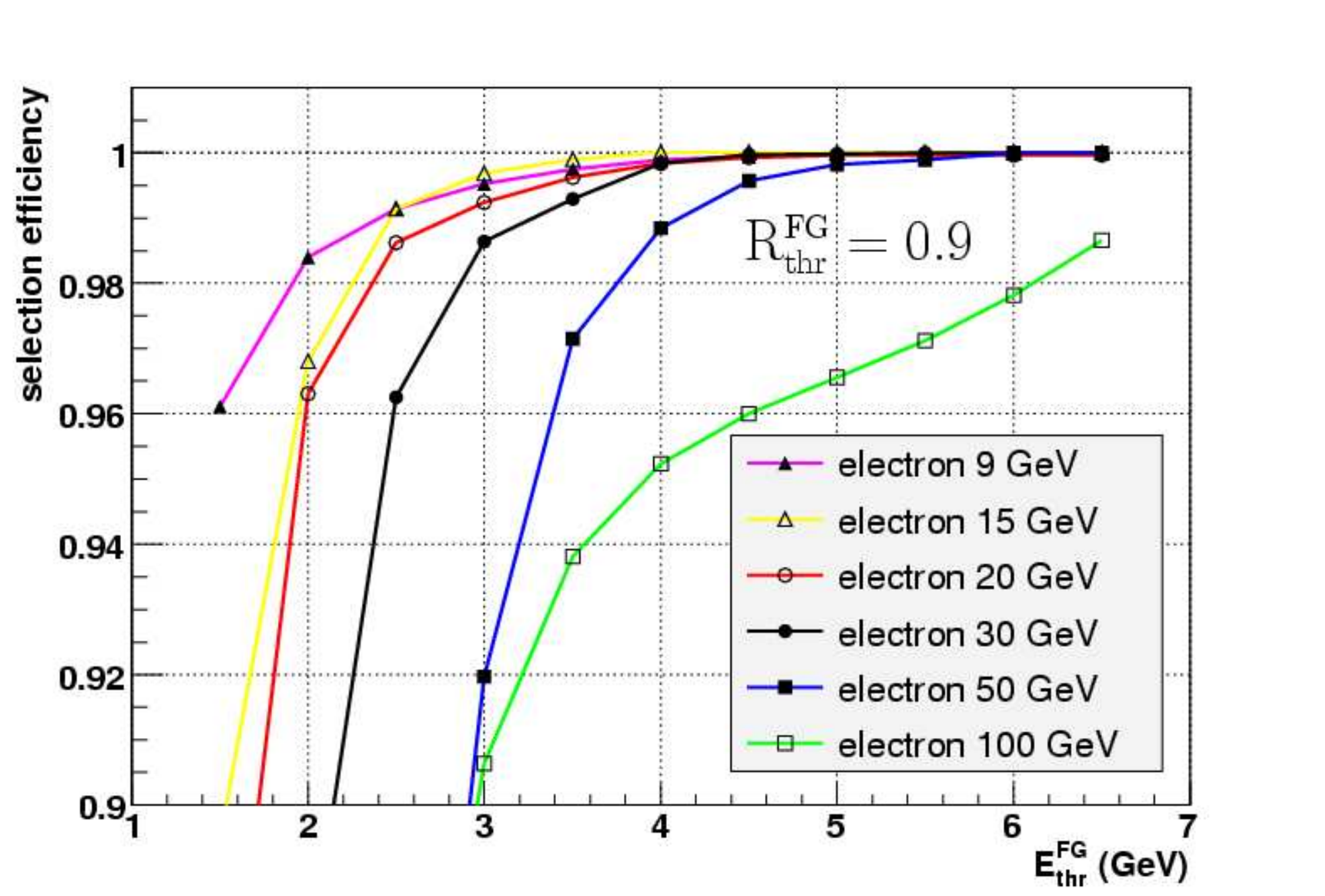}\tabularnewline
\end{tabular}\end{center}

\caption{Fraction of events that survive the FG veto for all eight neighbor
towers as a function of $E\mathrm{_{thr}^{FG}}$ for a fixed $R\mathrm{_{thr}^{FG}}$
of : 0.8 (left) and 0.9 (right), for different electron beam energies.\label{cap:figure5}}
\end{figure}

\begin{figure}[H]
\begin{minipage}[c]{0.65\textwidth}%
\begin{center}\includegraphics[%
  width=0.85\textwidth]{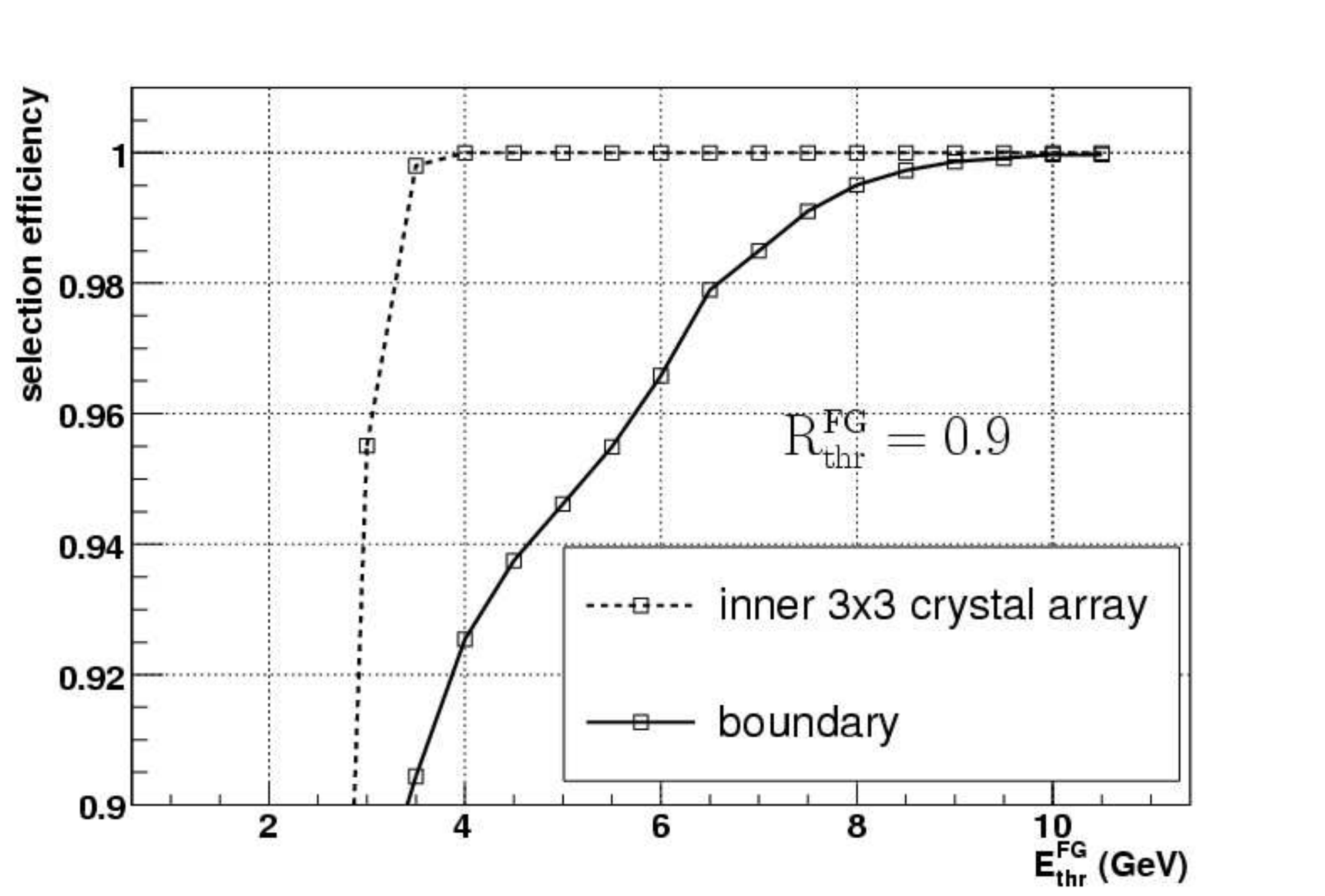}\end{center}\end{minipage}%
\hfill{}\begin{minipage}[c]{0.30\textwidth}%
\begin{flushleft}\includegraphics[%
  width=0.85\textwidth]{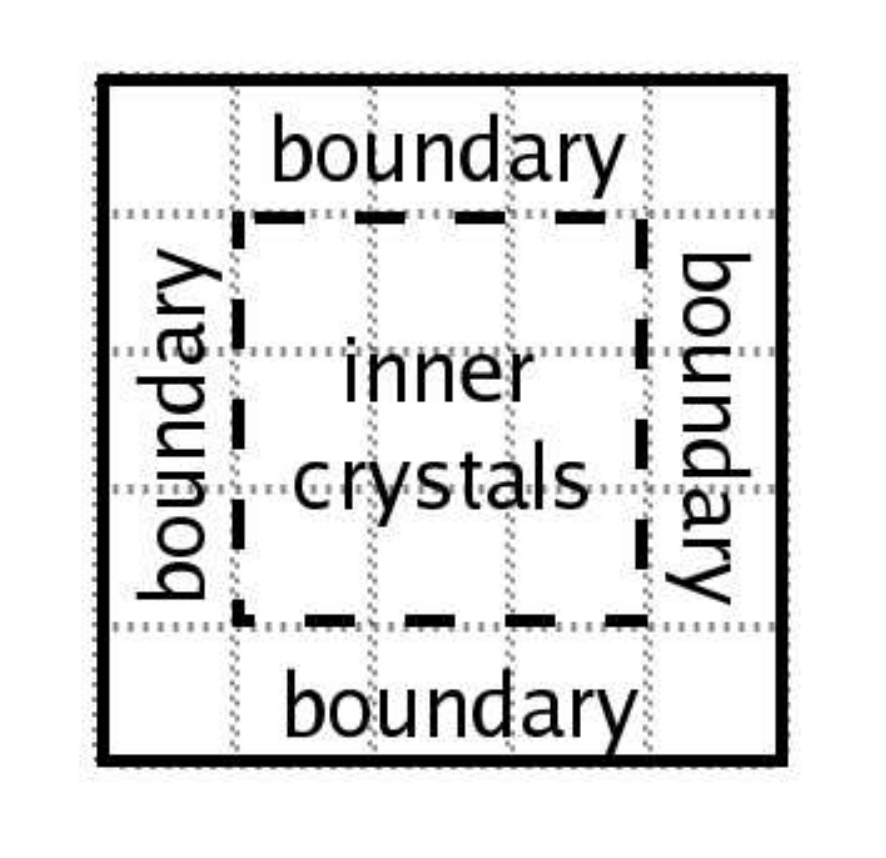}\end{flushleft}\end{minipage}%

\caption{Fraction of events in a 100 GeV electron beam that survive the FG
veto in all eight neighbor towers as a function of the tower energy
threshold when the most energetic crystal is located in the inner
$3\times3$ crystal array (dotted), and in the boundary crystals (solid).\label{cap:Fraction-of-events}}
\end{figure}

From the previous remarks, it is possible to reduce the rate of isolated
electrons which appear as non-isolated by increasing the energy threshold
of the trigger tower $(E\mathrm{_{thr}^{FG}}$) during the calculation
of FG veto bit. This threshold, however, depends on the energy of
the impinging electron. Due to the hardware design, it is not possible
to adjust $E\mathrm{_{thr}^{FG}}$ as a function of the electron energy.
Therefore, $E\mathrm{_{thr}^{FG}}$ needs to be optimized for the
energy range of electrons that are foreseen to be selected by a given
trigger stream. Figure \ref{cap:figure5} shows, for different
nominal beam energies, the fraction of events surviving the FG veto
for all eight neighbor towers as a function of $E\mathrm{_{thr}^{FG}}$.
Since the original distribution of the electron shower position relative
to the \emph{HIT} trigger tower boundaries is not uniform for different
beam energies, a simple event weighting technique was used, so that
the fraction of events hitting the inner $3\times3$ crystal array
was made constant for all energies. The selection efficiency on the
left plot was obtained using a threshold $R\mathrm{_{thr}^{FG}}=0.8$
, while for the right plot a value of 0.9 was used. For $R\mathrm{_{thr}^{FG}}=0.8$,
and discarding all towers with ECAL energy < 3 GeV, an efficiency
above 99\% is measured for all electron energies which were considered.
For $R\mathrm{_{thr}^{FG}}=0.9$, however, it is necessary to raise
the $E\mathrm{_{thr}^{FG}}$ to 5 GeV to achieve an efficiency above
99\% for electron energies up to \textasciitilde{} 50 GeV. The electrons
from test beam are supposed to be isolated for all energies, however
a non negligible fraction of 100 GeV electrons is classified as non-isolated
(Figure \ref{cap:figure5}, right) as the energy leakage to neighboring
towers becomes significant. The loss of isolation efficiency for electrons
with energy of 100 GeV is driven by events in which the most energetic
crystal is located near the boundary of the trigger tower. This is
illustrated in Figure \ref{cap:Fraction-of-events}, where the fraction
of 100 GeV electrons events that survive the FG veto for all eight
neighbor towers is shown as a function of $E\mathrm{_{thr}^{FG}}$
and electron shower position relative to the trigger tower. A value
of $R\mathrm{_{thr}^{FG}}=0.9$ is used. It is important to stress
that the electrons misclassified as non-isolated will still be recorded
in the non-isolated trigger stream, as long as the threshold on $E\mathrm{_{T}^{cand}}$
is low enough. This is valid under the assumption that the High Level
Trigger electron stream is seeded by both isolated and non-isolated
L1 objects. For low-luminosity ($~2\times10^{33}{\rm cm^{2}s^{-1}}$)
the threshold on $E\mathrm{_{T}^{cand}}$ is about 20 GeV for the
L1 single electron stream \cite{CMS_Note_2006-085}, while for high-luminosity
$(10^{34}{\rm cm^{2}s^{-1}})$ it is about 50 GeV, which may affect
the electron selection efficiency \cite{CMS_Note_2000-074}. Regarding
the threshold on the tower energy for the FG veto bit, a possible
side effect of its increase would be a loss of jet rejection power
for the electron stream. This effect, though, cannot be assessed,
as non-isolated hadron jet data are not yet available.

\section{Electron Selection Efficiency : Results\label{sec:Results-on-Electron}}

\begin{figure}[H]
\begin{center}\begin{tabular}{cc}
\includegraphics[%
  width=0.49\textwidth]{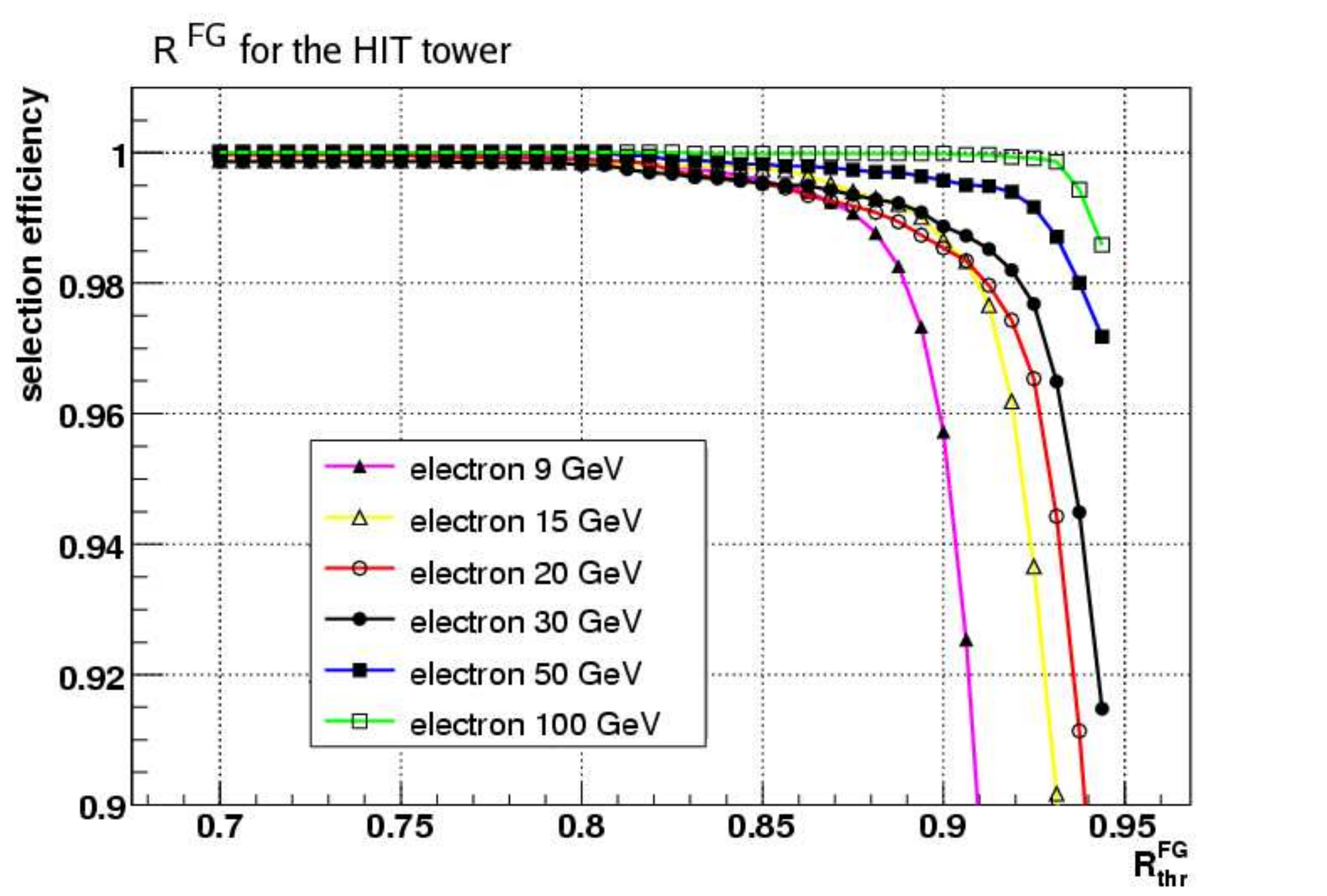}&
\includegraphics[%
  width=0.49\textwidth]{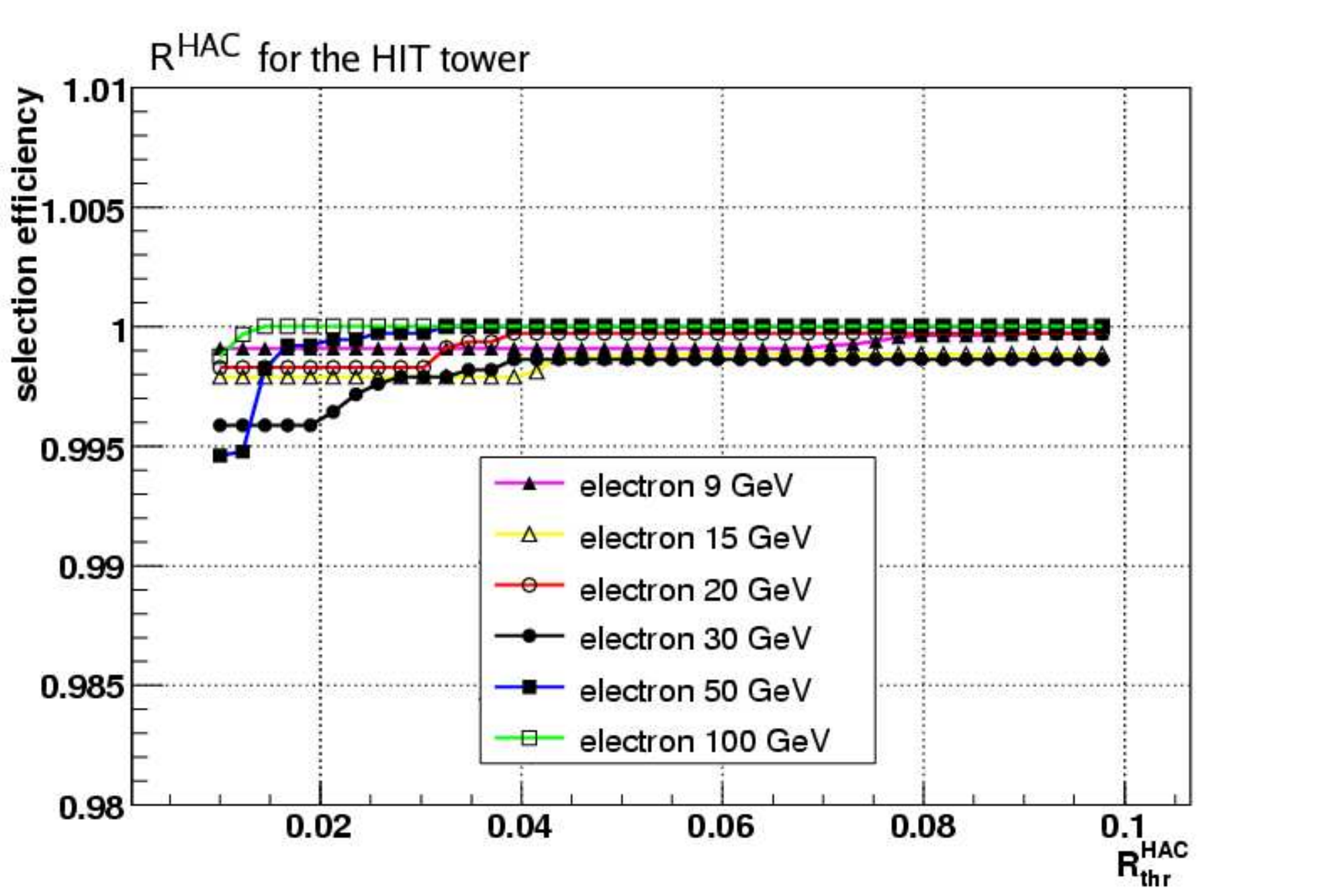}\tabularnewline
\includegraphics[%
  width=0.49\textwidth]{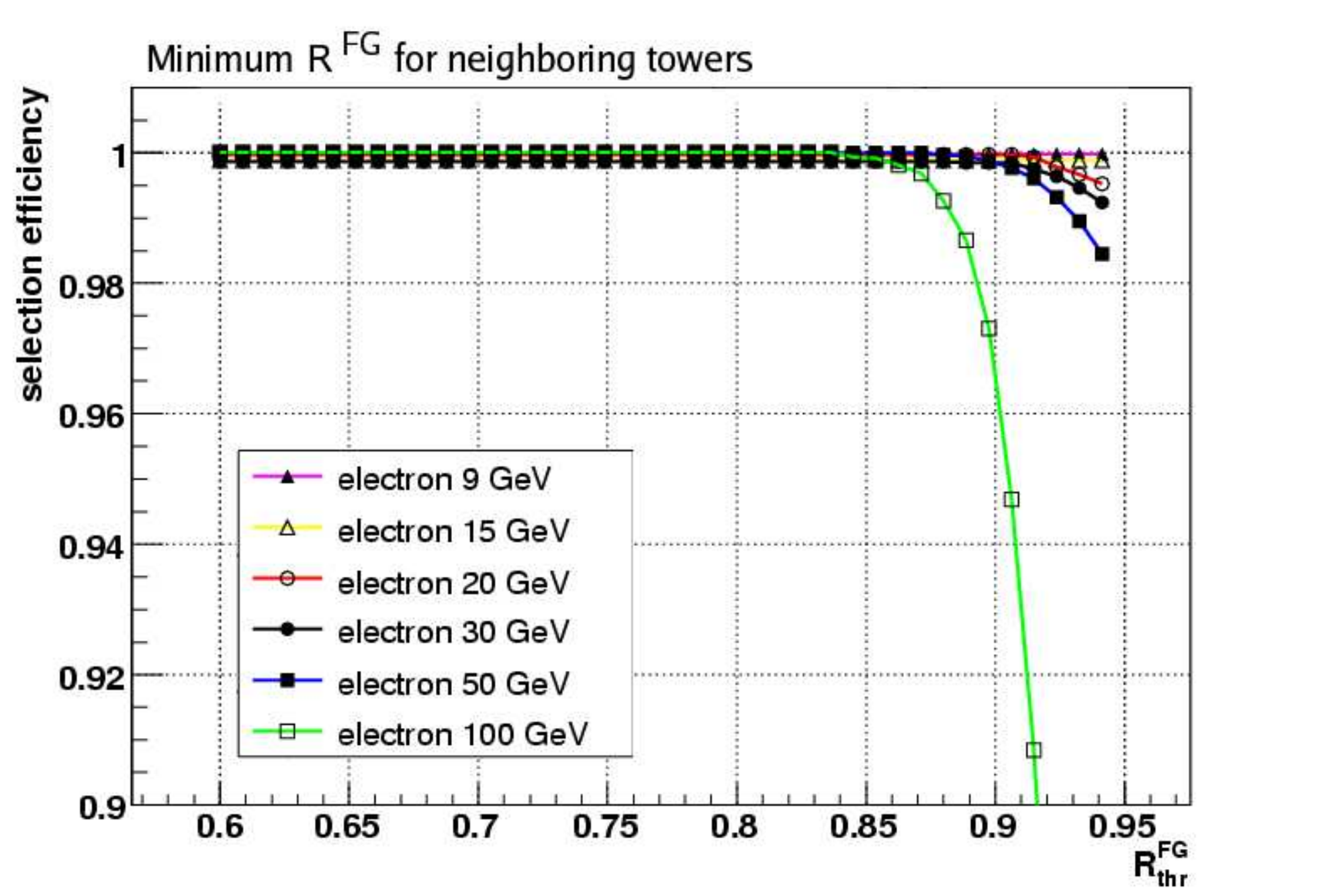}&
\includegraphics[%
  width=0.49\textwidth]{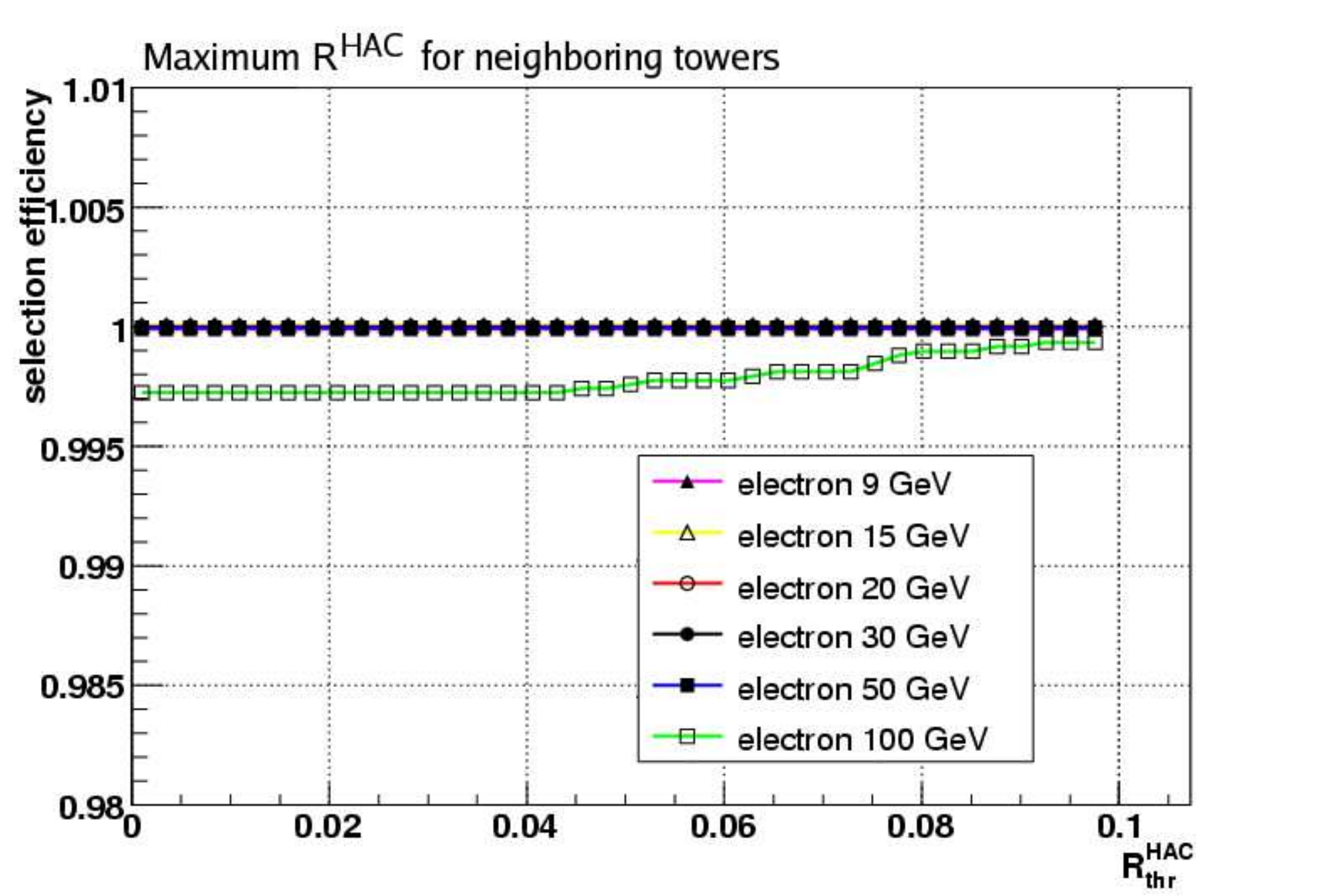}\tabularnewline
\includegraphics[%
  width=0.49\textwidth]{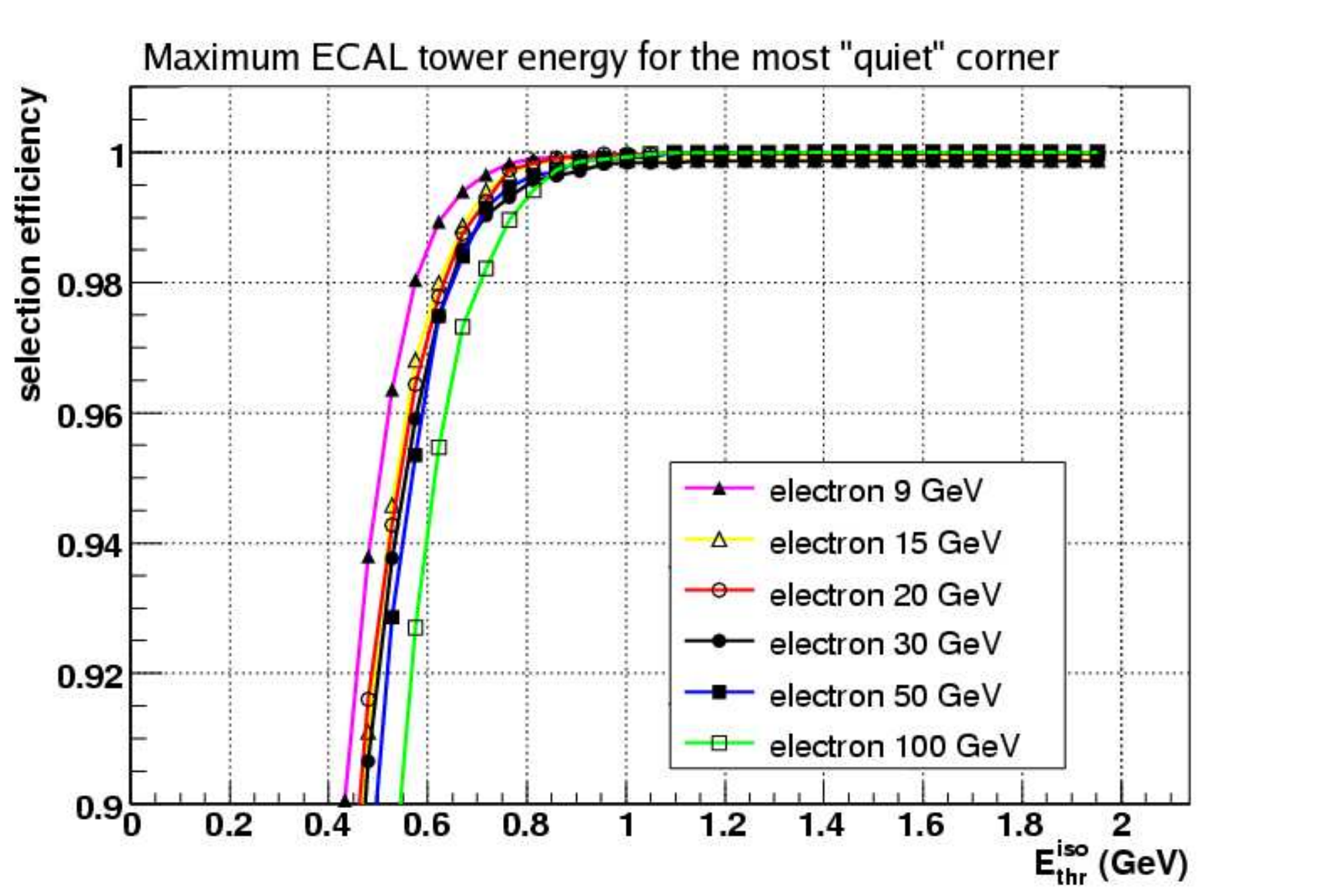}&
\includegraphics[%
  width=0.49\textwidth]{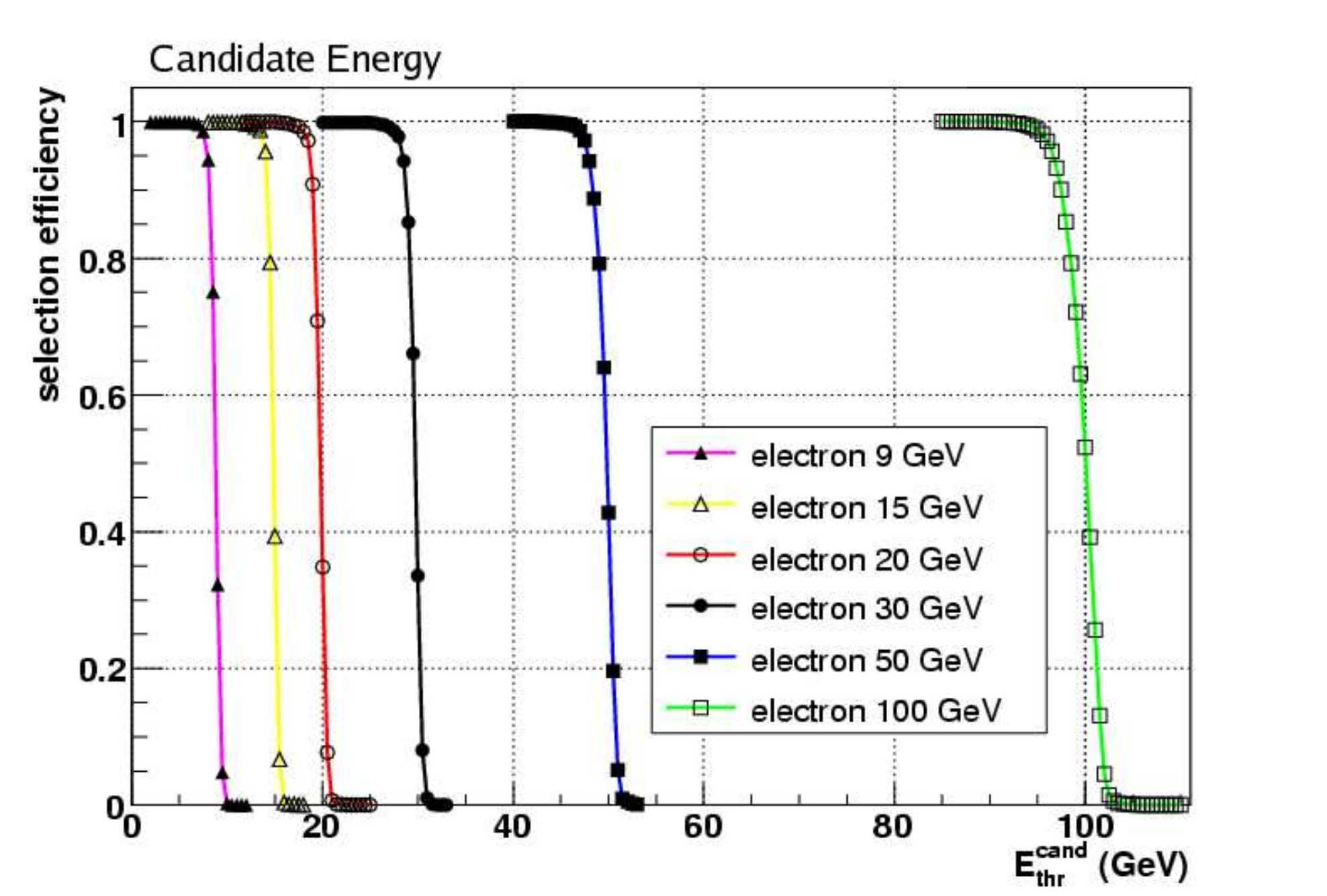}\tabularnewline
\end{tabular}\end{center}

\caption{Dependence of the electron selection efficiency on the threshold
used for each criterion of the L1 electron trigger algorithm.\label{cap:figure6}}
\end{figure}

The dependence of the electron efficiency for each selection criterion
used by the L1 algorithm is shown separately in Figure \ref{cap:figure6}
for all energy samples. The average selection efficiency for electrons
impinging on any of the $5\times5$ crystals of the \emph{HIT} trigger
tower is shown as a function of the threshold for a given selection
variable considered individually, regardless of other selection cuts.
The event weighting technique mentioned in the previous section was
also applied. The plots on the top row refer to the electron selection
efficiency of FG and HAC vetoes, relative to the trigger tower where
the electron hits (\emph{HIT} tower). Following the discussion presented
in Section \ref{sec:Impact-of-Combined}, the energy thresholds were
set at $E\mathrm{_{thr}^{FG}}=E\mathrm{_{thr}^{HAC}}=5{\rm GeV}$.
The plots in the middle row refer to the first isolation criteria.
For each event the neighbor trigger tower with the minimum FG ratio
and the one with the maximum HAC ratio was found. According to the
thresholds, it was determined if the event would satisfy the requirement.
The convolution of the efficiencies presented in the two plots of
the middle row gives the efficiency for the Combined Neighbor Veto.
The plot on the bottom left refers to the selection efficiency of
the ECAL isolation veto. The thresholds for all the previous discriminators
are programmable and can be reset for each run. The electron selection
efficiency is mostly sensitive to the FG veto requirement as it can
be seen in Figure \ref{cap:figure6}, top left and middle left. The
energy distribution of the electron candidate events is shown as an
efficiency curve as a function of $E\mathrm{_{thr}^{cand}}$ (Figure
\ref{cap:figure6}, bottom right) without taking into account the
effect of TPG digitization on energy resolution.

Test beam data was used in this study to validate the performance
of the electron trigger selection and determine the efficiency given
reference thresholds rather than to tune them. In Section \ref{sec:Impact-of-Combined},
an attempt was made to improve the electron efficiency by optimizing
$E\mathrm{_{thr}^{FG}}$. To illustrate the impact of $E\mathrm{_{thr}^{FG}}$
on the electron classification, results are presented for two values
of $E\mathrm{_{thr}^{FG}}$ : 3 and 5 GeV, the latter being the optimal
value for electrons energies up to \textasciitilde{} 50 GeV. All other
threshold values determined from previous simulation studies \cite{CMS_Note_1999-026,CMS_TN_96-021}
seemed reasonable and the reference values were used. The selection
cuts used in the study of the L1 algorithm efficiency are summarized
in Table \ref{cap:thresholds}. The results on electron selection
efficiency are presented in Table \ref{cap:results} for the isolated
and non-isolated stream. The efficiency results reported are lower
limits estimated assuming a binomial distribution and considering
one-sided intervals containing 95\% of the probability.

\begin{table}[H]

\begin{center}\begin{tabular}{|l|c|}
\hline 
\multicolumn{1}{|c|}{Criterion}&
Threshold\tabularnewline
\hline
\hline 
$E\mathrm{^{FG}}$\hfill{} (FG Tower Energy)&
3,$\;$5 GeV\tabularnewline
\hline
$E\mathrm{^{HAC}}$ \hfill{}(HAC Tower Energy)&
5 GeV\tabularnewline
\hline
$R\mathrm{^{FG}}$\hfill{}(FG ratio)&
0.9\tabularnewline
\hline
$R\mathrm{^{HAC}}$\hfill{} (HAC ratio)&
0.05\tabularnewline
\hline
$E\mathrm{^{iso}}$ \hfill{}(Ecal Isolation)&
1.5 GeV\tabularnewline
\hline
\end{tabular}\end{center}

\caption{Thresholds used to evaluate the overall L1 electron selection efficiency,
for each criterion.\label{cap:thresholds}}

\end{table}

\begin{table}[H]

\begin{center}\begin{tabular}{|c|c|c||c|c|}
\cline{2-5} 
\multicolumn{1}{c|}{}&
\multicolumn{4}{c|}{Electron selection efficiency lower limit (95\% CL)}\tabularnewline
\cline{2-5} 
\cline{2-3} \cline{4-5} 
\multicolumn{1}{c|}{}&
\multicolumn{2}{c||}{Non-isolated Stream}&
\multicolumn{2}{c|}{Isolated Stream}\tabularnewline
\cline{2-3} \cline{4-5} 
\hline 
\multicolumn{1}{|c|}{Energy }&
$E\mathrm{_{thr}^{FG}}$ = 3 GeV&
$E\mathrm{_{thr}^{FG}}$ = 5 GeV&
$E\mathrm{_{thr}^{FG}}$ = 3 GeV&
$E\mathrm{_{thr}^{FG}}$ = 5 GeV\tabularnewline
\hline
\hline 
9 GeV&
> 95\%&
> 95\%&
> 95\%&
> 95\%\tabularnewline
\hline
15 GeV&
> 98\%&
> 98\%&
> 98\%&
> 98\%\tabularnewline
\hline
20 GeV&
> 98\%&
> 98\%&
> 97\%&
> 98\%\tabularnewline
\hline
30 GeV&
> 99\%&
> 99\%&
> 97\%&
> 99\%\tabularnewline
\hline
50 GeV&
> 99\%&
> 99 \%&
> 91\%&
> 99\%\tabularnewline
\hline
100 GeV&
100\%&
100\%&
> 90\%&
> 96\%\tabularnewline
\hline
\end{tabular}\end{center}

\caption{Results for electron selection efficiency of L1 trigger algorithm,
under test beam conditions, for events passing shower profile cuts
(FG and HAC) for \emph{HIT} tower (non-isolated stream) and events
passing all cuts including isolation (isolated stream). Two values
are considered for $E\mathrm{_{thr}^{FG}}$: 3 and 5 GeV. The numbers
in the table are lower limits at 95\% confidence level for the selection
efficiency in \%.\label{cap:results}}

\end{table}

Electrons with energies from 15 GeV up to 100 GeV are selected in
the non-isolated stream with efficiency of 98--100\%. The lower trigger
efficiency for electrons with 9 GeV, 95\%, is due to lower $R\mathrm{^{FG}}$
values measured in the \emph{HIT} trigger tower, as shown in the top
left plot of Figure \ref{cap:figure6}. The selection efficiency of
the non-isolated stream is the same for $E\mathrm{_{thr}^{FG}}=3$
GeV and $E\mathrm{_{thr}^{FG}}=5$ GeV. For $E\mathrm{_{thr}^{FG}}=5$
GeV, electrons with nominal energies from 15 GeV up to 50 GeV are
classified as isolated with 98\% efficiency or more. For electrons
with an energy of 100 GeV, although all would be selected by either
the isolated or non-isolated stream, 4\% of them are classified as
non-isolated due to the energy leakage in the neighboring towers as
discussed in Section \ref{sec:Impact-of-Combined}. From inspection
of Figure \ref{cap:figure6} and from the results in Table \ref{cap:results}
for $E\mathrm{_{thr}^{FG}}=3$ GeV, it is clear that the misclassification
of these electrons is due to the Combined Neighbor veto.

\section{Results on Charged Hadron Rejection\label{sec:Results-on-Rejection}}

The optimization of the thresholds used for each criterion must also
consider the minimization of the trigger rate from fake QCD jet events.
The rejection of the L1 electron/photon algorithm for charged hadrons
was estimated using charged hadron beam data samples with energies
from 3 up to 100 GeV. The same selection cuts chosen for the electron
efficiency study were applied and are shown in Table \ref{cap:thresholds_rej}.
Additionally, the impact of the candidate energy requirement on the
hadron rejection was considered by setting a 10 GeV threshold. The
rejection for each cut applied individually is reported in Table \ref{cap:rejection per cut}
for all considered beam energies. The cuts that yield the greatest
rejection power are the candidate energy requirement, the FG and the
HAC veto for the \emph{HIT} tower. For low hadron energies ($\leq$
9 GeV) the fraction of events that are rejected by the FG and HAC
veto for the \emph{HIT} tower is low ( < 15\%) because of the $E\mathrm{_{thr}^{HAC}}$and
$E\mathrm{_{thr}^{FG}}$ ( = 5 GeV) values required and necessary
to ensure high isolation efficiencies for electrons with energies
up to 50 GeV.

\begin{table}[H]

\begin{center}\begin{tabular}{|l|c|}
\hline 
\multicolumn{1}{|c|}{Criterion}&
Threshold\tabularnewline
\hline
\hline 
$E\mathrm{^{FG}}$\hfill{} (FG Tower Energy)&
5 GeV\tabularnewline
\hline
$E\mathrm{^{HAC}}$ \hfill{}(HAC Tower Energy)&
5 GeV\tabularnewline
\hline
$R\mathrm{^{FG}}$\hfill{}(FG ratio)&
0.9\tabularnewline
\hline
$R\mathrm{^{HAC}}$\hfill{} (HAC ratio)&
0.05\tabularnewline
\hline
$E\mathrm{^{iso}}$ \hfill{}(Ecal Isolation)&
1.5 GeV\tabularnewline
\hline
$E\mathrm{^{cand}}$\hfill{} (Candidate Energy)&
10 GeV\tabularnewline
\hline
\end{tabular}\end{center}

\caption{Thresholds used to evaluate the L1 rejection of charged hadrons,
for each criterion.\label{cap:thresholds_rej}}

\end{table}

\begin{table}[H]

\begin{center}\begin{tabular}{|c|c|c|c|c|c|c|}
\cline{2-7} 
\multicolumn{1}{c|}{}&
\multicolumn{6}{c|}{Hadron rejection per cut}\tabularnewline
\hline 
Energy &
Candidate Energy&
HAC \emph{HIT}&
FG \emph{HIT}&
HAC Neigh.&
FG Neigh.&
Ecal Iso.\tabularnewline
\hline 
3 GeV&
100\%&
0\%&
0\%&
0\%&
0\%&
0\%\tabularnewline
\hline 
5 GeV&
100\%&
2\%&
2\%&
0\%&
0\%&
0\%\tabularnewline
\hline 
7 GeV&
100\%&
7\%&
3\%&
0\%&
0\%&
0\%\tabularnewline
\hline 
9 GeV&
99\%&
13\%&
7\%&
0\%&
0\%&
0\%\tabularnewline
\hline 
15 GeV&
95\%&
34\%&
24\%&
1\%&
0\%&
0\%\tabularnewline
\hline 
20 GeV&
83\%&
43\%&
28\%&
2\%&
0\%&
0\%\tabularnewline
\hline 
30 GeV&
63\%&
54\%&
28\%&
4\%&
0\%&
0\%\tabularnewline
\hline 
50 GeV&
48\%&
63\%&
26\%&
12\%&
0\%&
0\%\tabularnewline
\hline 
100 GeV&
43\%&
66\%&
19\%&
19\%&
1\%&
0\%\tabularnewline
\hline
\end{tabular}\end{center}

\caption{Results on rejection for charged hadrons at different energies. The
numbers in the table are the rejection in \% after each cut is applied
individually.\label{cap:rejection per cut}}

\end{table}

The overall L1 rejection for charged hadrons after all trigger cuts,
including the 10 GeV threshold on the candidate energy is presented
in Table \ref{cap:overall rejection} for the isolated and non-isolated
trigger streams. The rejection results reported are estimated assuming
a binomial distribution and considering symmetrical intervals containing
95\% of the probability. For all beam energies the fraction of events
rejected by the trigger cuts for the non-isolated trigger stream is
always larger than \textasciitilde{} 93\%. The additional event rejection
due to isolation cuts is less than 1\% with respect to the fraction
of events already rejected by the shower profile cuts (FG and HAC)
for the \emph{HIT} tower. 

\begin{table}[H]

\begin{center}\begin{tabular}{|c|c|c|}
\cline{2-3} 
\multicolumn{1}{c|}{}&
\multicolumn{2}{c|}{Hadron rejection intervals (95\% CL)}\tabularnewline
\hline 
Energy &
Non-isolated stream&
Isolated stream\tabularnewline
\hline 
3 GeV&
100\%&
100\%\tabularnewline
\hline 
5 GeV&
100\%&
100\%\tabularnewline
\hline 
7 GeV&
100\%&
100\%\tabularnewline
\hline 
9 GeV&
{[}99.8 , 99.9{]}\%&
{[}99.8 , 99.9{]}\%\tabularnewline
\hline 
15 GeV&
{[}94.6 , 98.3{]}\%&
{[}95.0 , 98,5{]}\%\tabularnewline
\hline 
20 GeV&
{[}93.4 , 94.2{]}\%&
{[}93.8 , 94.6{]}\%\tabularnewline
\hline 
30 GeV&
{[}93.8 , 94.9{]}\%&
{[}94.2 , 95.3{]}\%\tabularnewline
\hline 
50 GeV&
{[}96.2 , 96.6{]}\%&
{[}97.2 , 97.5{]}\%\tabularnewline
\hline 
100 GeV&
{[}97.7 , 98.1{]}\%&
{[}98.5 , 98.8{]}\%\tabularnewline
\hline
\end{tabular}\end{center}

\caption{Results on rejection for charged hadrons at different energies for
events failing any shower profile cut (FG or HAC) for \emph{HIT} tower
(non-isolated stream) and events failing any cut, including isolation
(isolated stream). The numbers in the table are symmetrical intervals
at 95\% confidence level of the overall hadron rejection for the non-isolated
and isolated electron/photon stream in \%, including a cut on candidate
energy of 10 GeV .\label{cap:overall rejection}}

\end{table}

\section{Conclusions\label{sec:Conclusions}}

The electron selection efficiency of the L1 trigger algorithm is measured
for electrons of the combined ECAL/HCAL test beam data. Results indicate
that identification efficiencies of 98--100\% can be achieved for
electron energies from 15 up to 100 GeV. The selection efficiency
for the isolated electron stream is limited by the energy leakage
to the neighboring trigger towers which induces spurious neighbor
veto bits. In order to reduce this effect and improve the efficiency
for isolated electrons, a minimum threshold of 5 GeV on the tower
energy can be set. Following this approach more than 99\% of isolated
electrons with energies up to 50 GeV are also considered isolated
according to the Fine Grain veto, while the efficiency for the non-isolated
stream remains unaffected. The overall selection efficiency for the
isolated electron stream was found to be above 98\% for electrons
with energies from 15 up to 50 GeV. Due to the energy leakage to neighboring
towers, approximately 4\% of the isolated electrons with 100 GeV are
considered as non-isolated. If a proper classification is necessary
for energies higher than 50 GeV either the Combined Neighbor veto
should not be applied or the threshold on the tower energy for Fine
Grain veto should be adjusted.

The fraction of charged hadrons with energies from 3 up to 100 GeV
rejected by the L1 electron trigger algorithm is estimated to be larger
than 93\%.

\acknowledgments{
The authors would like to express their gratitude to the 2006 Combined
ECAL/HCAL test beam community for providing detailed information on
the experimental setup, data access and particle identification methods.
We thank in particular S. Kunori, A. Askew, L. Berntzon J. Damgov
and F. Ratnikov. We also acknowledge fruitful discussions with S.
Dasu and W. Smith regarding the CMS L1 electron trigger algorithm.
}

\newpage

\appendix

\section{Pseudo-code for Fine Grain bit evaluation\label{sec:appendix A}}

\begin{itemize}
\item if (Tower Energy < $E{\rm {_{thr\_ low}^{FG}}}$) FG veto bit = 0
\item if ($E{\rm {_{thr\_ low}^{FG}}}$<Tower Energy <$E{\rm {_{thr\_ high}^{FG}}}$) 

\begin{itemize}
\item if ( FG ratio < $R{\rm {_{thr\_ low}^{FG}}}$ ) FG veto bit = 1
\item else FG veto bit = 0
\end{itemize}
\item if (Tower Energy > $E{\rm {_{thr\_ high}^{FG}}}$) 

\begin{itemize}
\item if (FG ratio < $R{\rm {_{thr\_ low}^{FG}}}$) FG veto bit =1
\item else FG veto bit = 0
\end{itemize}
\end{itemize}

\newpage

\bibliography{cms}

\end{document}